\documentclass[a4paper,10pt,twoside]{cpc-hepnp}

\usepackage{multicol}
\usepackage{graphicx}
\usepackage{booktabs}
\usepackage{amssymb,bm,mathrsfs,bbm,amscd}
\usepackage[tbtags]{amsmath}
\usepackage{lastpage}
\usepackage{epstopdf}
\usepackage{CJK}
\usepackage{multirow}
\usepackage{amsfonts,amssymb}
\usepackage{pgf,pgfarrows,pgfnodes,pgfautomata,pgfheaps}
\usepackage{mathrsfs}
\usepackage{verbatim}
\usepackage{array}
\newcommand{\PreserveBackslash}[1]{\let\temp=\\#1\let\\=\temp}
\newcolumntype{C}[1]{>{\PreserveBackslash\centering}p{#1}}
\newcolumntype{R}[1]{>{\PreserveBackslash\raggedleft}p{#1}}
\newcolumntype{L}[1]{>{\PreserveBackslash\raggedright}p{#1}}

\begin{document}
\begin{CJK*}{GB}{gbsn}

%\fancyhead[c]{\small Chinese Physics C~~~Vol. XX, No. X (XXXX)
%XXXXXX} \fancyfoot[C]{\small XXXXXX-\thepage}

%\footnotetext[0]{Received XX XXXX XXXX}

\title{Leading order relativistic hyperon-nucleon interactions in chiral effective field theory\thanks{Supported by the National Natural Science Foundation of China under Grants No. 11375024, No. 11522539, and No. 11375120, the China Postdoctoral Science Foundation under Grant No. 2016M600845 and No. 2017T100008, and the Fundamental Research Funds for the Central Universities. }}

\author{%
      LI Kai-Wen$^{1}$
, REN Xiu-Lei$^{2}$
, GENG Li-Sheng$^{1,3;1)}$\email{lisheng.geng@buaa.edu.cn}%
, LONG Bingwei$^{4}$
}
\maketitle

\address{%
$^1$ School of Physics and Nuclear Energy Engineering and International Research Center for Nuclei and Particles in the Cosmos, Beihang University, Beijing 100191, China\\
$^2$ School of Physics and State Key Laboratory of Nuclear Physics and Technology, Peking University, Beijing 100871, China\\
$^3$ Beijing Key Laboratory of Advanced Nuclear Materials and Physics, Beihang University, Beijing 100191, China\\
$^4$ Center for Theoretical Physics, Department of Physics, Sichuan University, 29 Wang-Jiang Road, Chengdu, Sichuan 610064, China\\
}

\begin{abstract}
We apply a recently proposed covariant power counting in nucleon-nucleon interactions to study strangeness $S=-1$ $\Lambda N-\Sigma N$ interactions in chiral effective field theory. At leading order, Lorentz invariance  introduces 12 low energy constants, in contrast to the heavy baryon approach, where only five appear. The Kadyshevsky equation is adopted to resum the potential in order to account for the non-perturbative nature of hyperon-nucleon interactions. A fit to the $36$ hyperon-nucleon scattering data points yields  $\chi^2\simeq 16$, which is comparable with the sophisticated phenomenological models and the next-to-leading order heavy baryon approach. However, one cannot achieve a simultaneous description of the nucleon-nucleon phase shifts and strangeness $S=-1$ hyperon-nucleon scattering data at leading order.
%The relativistic interactions might provide important inputs to relativistic hypernuclear structure studies.
\end{abstract}



%\footnotetext[0]{\hspace*{-3mm}\raisebox{0.3ex}{$\scriptstyle\copyright$}2013
%Chinese Physical Society and the Institute of High Energy Physics
%of the Chinese Academy of Sciences and the Institute
%of Modern Physics of the Chinese Academy of Sciences and IOP Publishing Ltd}%

%\begin{multicols}{2}

\section{Introduction}
\label{Section:Intro}

Since the quantum number \textit{strangeness} was introduced \cite{GellMann:1953zza,Nakano:1953zz} and the first observation of $\Lambda$ hypernuclei \cite{Danysz:1953xy} in 1953, strangeness nuclear physics has always been at the frontier of  experimental and theoretical nuclear physics. In recent years, open questions such as  charge symmetry breaking in $A=4$ $\Lambda$-hypernuclei \cite{Bedjidian:1979qh} and the existence of the H-dibaryon \cite{Jaffe:1976yi} have attracted a lot of attention~\cite{Nogga:2001ef,Gazda:2015qyt,Beane:2010hg,Inoue:2010es,Haidenbauer:2011ah,Yamaguchi:2016kxa}. In facilities like JLab, J-PARC, KEK, MAMI, and COSY, many important studies are being pursued, e.g., the level spectra and decay properties of $\Lambda$, double $\Lambda$ and $\Xi$ hypernuclei \cite{Gogami:2015tvu,Esser:2015trs,Yamamoto:2015avw,Ahn:2013poa,Nakazawa:2015joa}, $\Sigma p$ scattering \cite{jparc}, and final state interactions in production reactions, such as $\vec{p}p\rightarrow K^+ \Lambda p$ \cite{Hauenstein:2016zys}, which can provide information on the $\Lambda N$ scattering lengths. Meanwhile, theoretical few- and many-body calculations of hypernuclei have made steady progress, see, e.g., Refs.~\cite{Hiyama:2014cua,Zhou:2007zze}.  One particularly interesting  ongoing issue is about the role of hyperons in the cores of neutron stars, known as the \textit{hyperon puzzle}: nuclear many-body calculations incorporating hyperon degrees of freedom~\cite{Massot:2012pf,Schulze:2011zza,Hu:2013tma,Miyatsu:2013hea,Mallick:2012wb} have difficulties in obtaining a two-solar mass neutron star that was recently observed~\cite{Demorest:2010bx,Antoniadis:2013pzd}.

As the most important theoretical input for few- and many-body calculations, baryon-baryon interactions play an indispensable role in studies of hypernuclear physics. Although many efforts have been made to derive them, previous theoretical investigations were mainly based on phenomenological meson-exchange models \cite{Nagels:1976xq,Maessen:1989sx,Rijken:1998yy,Rijken:2006ep,Nagels:2015lfa,Holzenkamp:1989tq,Reuber:1993ip,Haidenbauer:2005zh} and quark models \cite{Straub:1988gj,Straub:1990de,Zhang:1994pp,Zhang:1997ny,Ping:1998si,Fujiwara:1995fx,Fujiwara:2006yh}. In the past two decades, two breakthroughs have occurred in constructing model-independent baryon-baryon interactions. Both of them are closely related to quantum chromodynamics (QCD), the underlying theory of strong interactions. One breakthrough is lattice QCD simulations \cite{Beane:2006gf,Nemura:2008sp,Beane:2009py,Inoue:2010hs,Beane:2011iw,Sasaki:2015ifa}, which provide an \textit{ab initio} numerical solution to QCD from first principles. With ever-growing  computing power and evolving numerical algorithms, lattice QCD simulations are approaching the physical world \cite{Doi:2015oha,Doi:2015uvd}, thus providing us with more information and constraints on baryon-baryon interactions.
The other is chiral effective field theory ($\chi$EFT), which has achieved great successes in nucleon-nucleon ($NN$) interactions \cite{Bedaque:2002mn,Epelbaum:2008ga,Machleidt:2011zz} following Weinberg's proposal~\cite{Weinberg:1990rz,Weinberg:1991um}. The latter approach has been generalized to antinucleon-nucleon \cite{Kang:2013uia}, hyperon-nucleon ($YN$) \cite{Polinder:2006zh,Haidenbauer:2007ra,Haidenbauer:2013oca} and multi-strangeness systems \cite{Polinder:2007mp,Haidenbauer:2009qn,Haidenbauer:2015zqb}. The main advantage of $\chi$EFT is that by using a power counting scheme, one can improve calculations systematically by going to higher orders in powers of external momenta and light quark masses, and estimating the uncertainties of any given order. Furthermore, three- and four-body forces automatically arise as we push through the hierarchy of chiral forces.

However,
the Weinberg approach for baryon-baryon interactions, denoted as the heavy baryon (HB) approach, is based on a non-relativistic formalism.
It is sensitive to ultraviolet cutoffs, that is, renormalization group invariance is violated, risking severe model dependence of short-range physics. Various opinions on this issue can be found in Refs.~\cite{Lepage:1997cs,Birse:2005um,Nogga:2005hy,Epelbaum:2006pt,Long:2007vp,Yang:2009pn,Valderrama:2009ei,Long:2011xw}. In two recent papers, Epelbaum and Gegelia have proposed a new approach  (referred to as the EG approach in the present paper) to $NN$ scattering in $\chi$EFT \cite{Epelbaum:2012ua,Epelbaum:2015sha}, where the relativistic effects are partially retained. At leading order (LO), the potential remains unchanged but the scattering equation changes to the Kadyshevsky equation, compared to the Lippmann-Schwinger equation with nonrelativistic nucleon propagators in the HB approach. Although this turned out to describe the Nijmegen partial wave analysis \cite{Stoks:1993tb} well, a higher order contact term is still needed in the $^3P_0$ partial wave to achieve renormalization group invariance. We applied the EG approach to the strangeness $S=-1$ $YN$ system \cite{Li:2016paq} and found that the best description of the experimental data is quantitatively similar to that of the HB approach, and that cutoff dependence is mitigated but not removed.

Partly motivated by the successes of covariant $\chi$EFT in the one-baryon and heavy-light systems \cite{Geng:2008mf,Geng:2009ik,Geng:2011wq,Ren:2012aj,Ren:2014vea,Geng:2010vw,Geng:2010df,Altenbuchinger:2011qn,Geng:2013xn},  a new covariant power counting is explored in Ref.~\cite{Ren:2016jna} to study $NN$ chiral interactions. The covariant treatment of baryons maintains all the symmetries and analyticities, and, at LO, it results in  a description of the $NN$ phase shifts  similar  to  that of the next-to-leading order (NLO) HB scheme.
In the present study, we apply this scheme to $YN$ scattering with strangeness $S=-1$, where more particle channels and less experimental data should be dealt with.

\section{Formalism\label{sec_formalism}}

\subsection{Covariant power counting}
First, we explain in some detail the covariant power counting scheme proposed in  Ref.~\cite{Ren:2016jna}~\footnote{See, also, Refs.\cite{Djukanovic:2007zz,Girlanda:2010ya} for early attempts.}.
Unlike the meson-meson and meson-baryon sectors, such a power counting in the baryon-baryon sector is not yet systematically formulated beyond leading order. In particular, relativistic contact baryon-baryon interactions should be treated carefully, see, e.g. Ref.~\cite{Girlanda:2010ya}.
%, because interactions involving meson-baryon vertices can be derived rather straightforwardly using the covariant baryon chiral perturbation theory.
In the covariant scheme, one takes the full Dirac spinors for the baryon fields and uses partial derivatives on the baryon/meson fields and meson mass insertions to increase the chiral order.

The perturbative expansion is consistent with conventional $\chi$EFT, in which the scattering amplitude is
expanded in terms of a small quantity over a large quantity. The former could be the meson momentum or mass, or the baryon three-momenta, and
the latter could be the $\rho$ meson mass or the nucleon mass or the chiral symmetry breaking scale.
%The effective potentials are expanded according to
%\begin{equation}
%  V_{\textrm{eff}} = V_{\textrm{eff}}(Q,g,\nu) = \sum_\nu Q^\nu \mathcal{V}_\nu(Q/\nu,g),
%\end{equation}
In Ref.~\cite{Ren:2016jna}, naive dimensional analysis is used to determine the chiral order $\nu$,
\begin{equation}
  \nu = 2 - \frac{1}{2}B + 2L + \sum_iv_i\Delta_i,~~~~~~\Delta_i=d_i+\frac{1}{2}b_i-2,
\end{equation}
where $B$ denotes the number of external baryons, $L$ is the number of Goldstone boson loops and $v_i$ is the number of vertices with dimension $\Delta_i$. For a vertex with dimension $\Delta_i$, $d_i$ is the number of derivatives or Goldstone boson masses, and $b_i$ is the number of internal baryon lines.

At leading order, there are no derivatives or pseudoscalar meson mass insertions. Therefore, the complete structures are determined by the Clifford algebra ($\Gamma_i$), namely the five Lorentz structures shown in the following section. These five structures
have been derived in a number of early studies in the nucleon-nucleon sector~\cite{Polinder:2006zh,Djukanovic:2007zz,Girlanda:2010ya}. Some authors consider the terms involving only $\gamma_5$ as higher order because they connect the large and small components of the Dirac spinors~\cite{Petschauer:2013uua}. In our present
case, we do not expand the Dirac spinors and therefore retain them.

\subsection{Leading order baryon-baryon interactions}

\begin{figure}
  \centering
  % Requires \usepackage{graphicx}
  \includegraphics[width=0.12\textwidth]{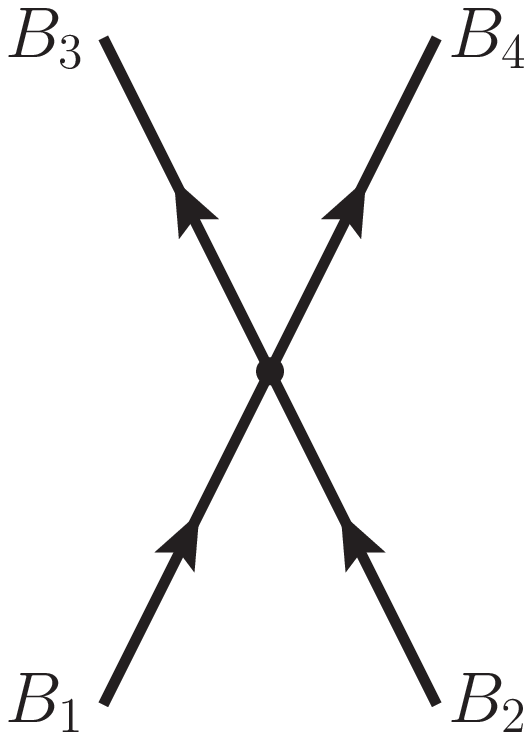} \qquad \qquad
  \includegraphics[width=0.12\textwidth]{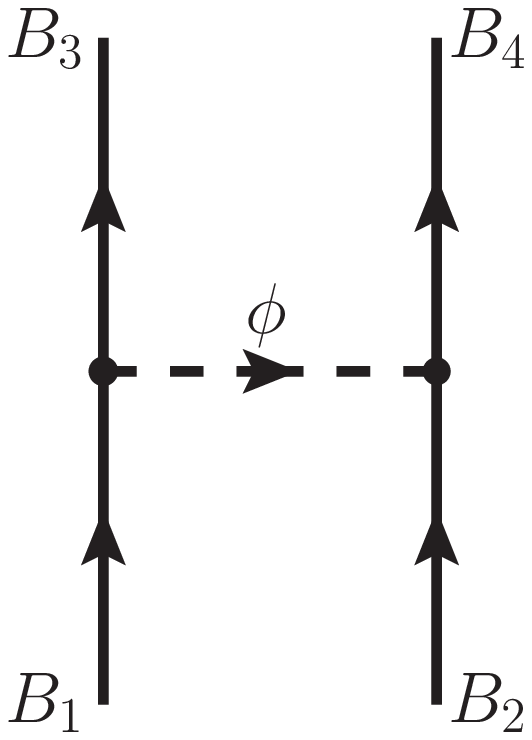}
  \caption{Non-derivative four-baryon contact terms and OPME at LO. The solid lines denote incoming and outgoing baryons ($B_{1,2,3,4}$), and the dashed line
  denotes the exchanged pseudoscalar meson $\phi$.}\label{CTOME}
\end{figure}

In covariant power counting~\cite{Ren:2016jna}, the full baryon spinor is retained to maintain Lorentz invariance
\begin{equation}\label{ub}
  u_B(\mbox{\boldmath $p$}, s)= N_p
  \left(
  \begin{array}{c}
    1 \\
    \frac{\mbox{\boldmath $\sigma$}\cdot \mbox{\boldmath $p$}}{E_p+M_B}
  \end{array}\right)
  \chi_s,
  ~~~~
  N_p=\sqrt{\frac{E_p+M_B}{2M_B}},
\end{equation}
where $E_p=\sqrt{\mbox{\boldmath $p$}^2+M_B^2}$, while a non-relativistic reduction of $u_B$ is employed in the HB approach.
The LO baryon-baryon interactions include non-derivative four-baryon contact terms (CT) and one-pseudoscalar-meson exchange (OPME) potentials, as shown in Fig.~\ref{CTOME},
\begin{equation}
  V_\mathrm{LO} = V_\mathrm{CT} + V_\mathrm{OPME}.
\end{equation}

\subsubsection{Four-baryon contact terms}
\begin{figure}
  \centering
  % Requires \usepackage{graphicx}
  \includegraphics[width=0.15\textwidth]{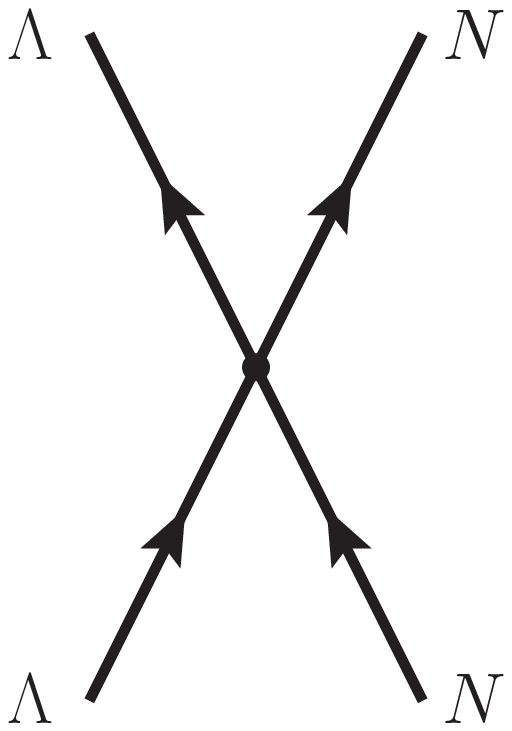}~~
  \includegraphics[width=0.15\textwidth]{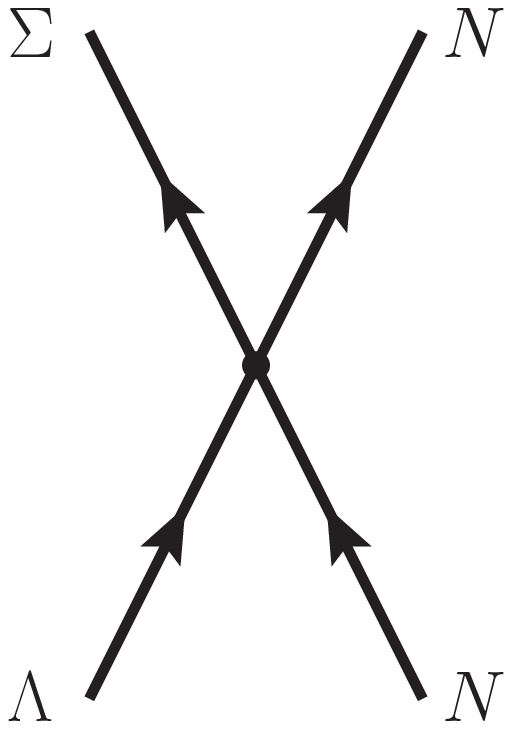}~~
  \includegraphics[width=0.15\textwidth]{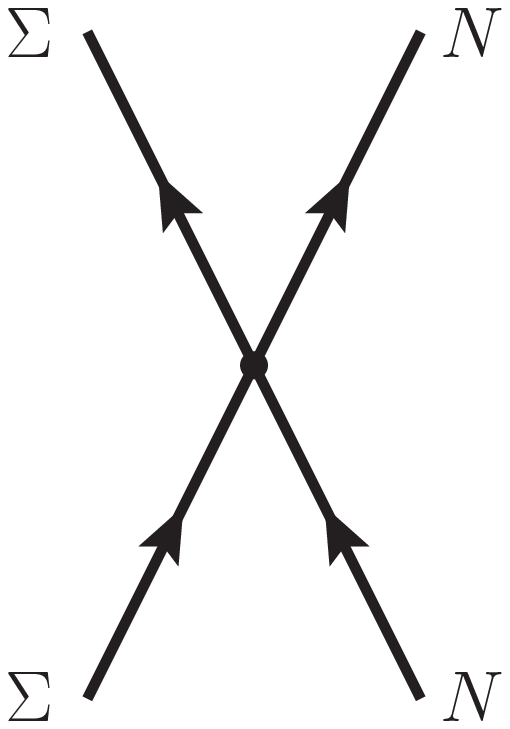}
  \caption{The non-derivative four baryon contact terms in the $\Lambda N-\Sigma N$ system.}\label{CT3}
\end{figure}

The Lagrangian term for non-derivative four-baryon contact interactions~\cite{Polinder:2006zh} is
\begin{align}\label{CT}
  \mathcal{L}_{\textrm{CT}} = \sum_{i=1}^5\left[\frac{\tilde C_i^1}{2}~\textrm{tr}\left(\bar B_1 \bar B_2 (\Gamma_i B)_2 (\Gamma_i B)_1\right)
  + \frac{\tilde C_i^2}{2}~\textrm{tr}\left(\bar B_1 (\Gamma_i B)_1 \bar B_2 (\Gamma_i B)_2\right)
  + \frac{\tilde C_i^3}{2}~\textrm{tr}\left(\bar B_1 (\Gamma_i B)_1\right)\textrm{tr}\left( \bar B_2 (\Gamma_i B)_2\right)\right],
\end{align}
where $\mathrm{tr}$ indicates the trace in flavor space ($u$, $d$, and $s$). Only baryon fields with the same subscript, 1 or 2, are grouped to form a Lorentz-covariant bilinear. $\Gamma_i$ are the elements of the Clifford algebra,
\begin{equation}\label{CA}
  \Gamma_1=1\, ,~~~~\Gamma_2=\gamma^\mu\, ,~~~~\Gamma_3=\sigma^{\mu\nu}\, ,
  ~~~~\Gamma_4=\gamma^{\mu}\gamma_5\, ,~~~~\Gamma_5=\gamma_5\, ;
\end{equation}
and $\tilde C_i^m$ ($m=1,2,3$) are the LECs corresponding to independent four-baryon operators. The ground-state octet baryons are collected in the $3\times3$ traceless matrix:
\begin{equation}\label{BMatrix}
  B =
  \left(
   \begin{array}{ccc}
    \frac{\Sigma^0}{\sqrt{2}}+\frac{\Lambda}{\sqrt{6}} & \Sigma^+ & p \\
    \Sigma^- & -\frac{\Sigma^0}{\sqrt{2}}+\frac{\Lambda}{\sqrt{6}} & n\\
    \Xi^- & \Xi^0 & -\frac{2\Lambda}{\sqrt{6}}
  \end{array}
  \right)\, .
\end{equation}
The Pauli exclusion principle applies; therefore, the two-baryon wave function is antisymmetric with respect to angular momentum $L$, spin $S$ and flavor. The flavor symmetric and flavor antisymmetric interactions are treated differently by using Fierz rearrangements, as has been done in Ref.~\cite{Polinder:2006zh}. The resulting Lagrangians for strangeness $S=-1$ $YN$ system in the isospin basis are shown in the following, corresponding to the three Feynman diagrams shown in Fig.~\ref{CT3}.
\vspace{3mm}
\par\noindent
$\bullet$~~The Lagrangians for the isospin $I=1/2$ $\Lambda N \rightarrow \Lambda N$ reaction are:
\begin{align}\label{lala1}
  & \mathcal{L}^{\Lambda\Lambda}_{1/2,\textrm{FS}} =
  \left(\frac{1}{6}C^{1}_i
  + \frac{5}{3}C^{2}_i
  + 2C^{3}_i\right)\left(\bar\Lambda \Gamma_i \Lambda\right)\left(\bar N \Gamma_i N\right)
  \equiv C^{\Lambda\Lambda}_{i,1/2,\textrm{FS}}\left(\bar\Lambda \Gamma_i \Lambda\right)\left(\bar N \Gamma_i N\right), \\
  & \mathcal{L}^{\Lambda\Lambda}_{1/2,\textrm{FA}} =
  \left(\frac{3}{2}C^{1}_i
  + C^{2}_i
  + 2C^{3}_i\right)\left(\bar\Lambda \Gamma_i \Lambda\right)\left(\bar N \Gamma_i N\right)
  \equiv C^{\Lambda\Lambda}_{i,1/2,\textrm{FA}}\left(\bar\Lambda \Gamma_i \Lambda\right)\left(\bar N \Gamma_i N\right),
\end{align}
where the subscripts FS and FA are short for flavor symmetric (e.g., $^1S_0$, $^3P_{0,1,2}\ldots$) and flavor antisymmetric (e.g., $^3S_1$, $^1P_1\ldots$), respectively.
\vspace{3mm}
\par\noindent
$\bullet$~~The Lagrangians for the isospin $I=3/2$ $\Sigma N \rightarrow \Sigma N$ reaction are:
\begin{align}
  & \mathcal{L}^{\Sigma\Sigma}_{3/2,\textrm{FS}} =
    2\left(C^{2}_i
  + C^{3}_i\right)\left(\bar\Sigma \Gamma_i \Sigma\right)\left(\bar N \Gamma_i N\right)
  \equiv C^{\Sigma\Sigma}_{i,3/2,\textrm{FS}}\left(\bar\Sigma \Gamma_i \Sigma\right)\left(\bar N \Gamma_i N\right) , \\
  & \mathcal{L}^{\Sigma\Sigma}_{3/2,\textrm{FA}} =
  - 2\left(C^{2}_i
  - C^{3}_i\right)\left(\bar\Sigma \Gamma_i \Sigma\right)\left(\bar N \Gamma_i N\right)
  \equiv C^{\Sigma\Sigma}_{i,3/2,\textrm{FA}}\left(\bar\Sigma \Gamma_i \Sigma\right)\left(\bar N \Gamma_i N\right).
\end{align}

\vspace{3mm}
\par\noindent
$\bullet$~~The Lagrangians for the isospin $I=1/2$ $\Sigma N \rightarrow \Sigma N$ reaction are:
\begin{align}
  & \mathcal{L}^{\Sigma\Sigma}_{1/2,\textrm{FS}} =
  \left(\frac{3}{2}C^{1}_i
  - C^{2}_i
  + 2C^{3}_i\right)\left(\bar\Sigma \Gamma_i \Sigma\right)\left(\bar N \Gamma_i N\right)
  \equiv C^{\Sigma\Sigma}_{i,1/2,\textrm{FS}}\left(\bar\Sigma \Gamma_i \Sigma\right)\left(\bar N \Gamma_i N\right)\notag\\
  &\phantom{\mathcal{L}^{\Sigma\Sigma}_{1/2,\textrm{FS}}}
  =\left(9C^{\Lambda\Lambda}_{i,1/2,\textrm{FS}}-8C^{\Sigma\Sigma}_{i,1/2,\textrm{FS}}\right)\left(\bar\Sigma \Gamma_i \Sigma\right)\left(\bar N \Gamma_i N\right),\label{sisi1}\\
  & \mathcal{L}^{\Sigma\Sigma}_{1/2,\textrm{FA}} =
  \left(\frac{3}{2}C^{1}_i
  + C^{2}_i
  + 2C^{3}_i\right)\left(\bar\Sigma \Gamma_i \Sigma\right)\left(\bar N \Gamma_i N\right)
  \equiv C^{\Sigma\Sigma}_{i,1/2,\textrm{FA}}\left(\bar\Sigma \Gamma_i \Sigma\right)\left(\bar N \Gamma_i N\right)\notag\\
  &\phantom{\mathcal{L}^{\Sigma\Sigma}_{1/2,\textrm{FA}}}
  = C^{\Lambda\Lambda}_{i,1/2,\textrm{FA}}\left(\bar\Sigma \Gamma_i \Sigma\right)\left(\bar N \Gamma_i N\right).\label{sisi2}
\end{align}

\vspace{3mm}
\par\noindent
$\bullet$~~The Lagrangians for the isospin $I=1/2$ $\Lambda N \rightarrow \Sigma N$ reaction are:
\begin{align}
  & \mathcal{L}^{\Lambda\Sigma}_{1/2,\textrm{FS}} =
  \left(\frac{1}{2}C^{1}_i
  - C^{2}_i\right)\left(\bar\Lambda \Gamma_i \Sigma\right)\left(\bar N \Gamma_i N\right)
  \equiv C^{\Lambda\Sigma}_{i,1/2,\textrm{FS}}\left(\bar\Lambda \Gamma_i \Sigma\right)\left(\bar N \Gamma_i N\right)\notag\\
  &\phantom{\mathcal{L}^{\Lambda\Sigma}_{1/2,\textrm{FS}}}
  =3\left(C^{\Lambda\Lambda}_{i,1/2,\textrm{FS}}-C^{\Sigma\Sigma}_{i,1/2,\textrm{FS}}\right)\left(\bar\Lambda \Gamma_i \Sigma\right)\left(\bar N \Gamma_i N\right),\label{lasi1}\\
  & \mathcal{L}^{\Lambda\Sigma}_{1/2,\textrm{FA}} =
  -\left(\frac{3}{2}C^{1}_i
  - C^{2}_i\right)\left(\bar\Lambda \Gamma_i \Sigma\right)\left(\bar N \Gamma_i N\right)
  \equiv C^{\Lambda\Sigma}_{i,1/2,\textrm{FA}}\left(\bar\Lambda \Gamma_i \Sigma\right)\left(\bar N \Gamma_i N\right).\label{lasi2}
\end{align}

\noindent
The superscript $YY'$ denotes the hyperons in the reaction of $YN \rightarrow Y'N$. Strict SU(3) symmetry is imposed, as shown in the second line of Eqs.~(\ref{sisi1}-\ref{lasi1}). Note that the LECs $C^{1,2,3}_i$ here are different from those in Eq.~(\ref{CT}) due to the application of Fierz rearrangement~\cite{Polinder:2006zh}. The potentials of the contact terms are derived from Eqs.~(\ref{lala1}-\ref{lasi2}), which can be symbolically written as
\begin{align}\label{ctsym}
  V^{YY'}_{\textrm{CT}} = C_i^{YY'}\left(\bar u_3 \Gamma_i u_1\right)\left(\bar u_4 \Gamma_i u_2\right) ,
\end{align}
where $C_i^{YY'}$ could be $C^{\Lambda\Lambda}_{i,1/2,\textrm{FS}}$, $C^{\Lambda\Lambda}_{i,1/2,\textrm{FA}}$, $C^{\Sigma\Sigma}_{i,1/2,\textrm{FS}}$, $C^{\Sigma\Sigma}_{i,1/2,\textrm{FA}}$ and $C^{\Lambda\Sigma}_{i,1/2,\textrm{FA}}$. They are first calculated in the helicity basis and then transformed to the $\left|LSJ\right>$ basis~\cite{Holzenkamp:1989tq}. We found that they contribute to all partial waves that have total angular momentum $J\leq 1$ (except for the $^1P_1-^3P_1$ mixing). We choose the LECs in $^1S_0$, $^3S_1$ and $^3P_1$ to be independent\footnote{The other choice is to take those in $^1S_0$, $^3S_1$ and $^3P_0$ partial waves.}, which is consistent with the $NN$ interactions \cite{Ren:2016jna}. The partial wave projected potentials are
%\footnote{Note that the appearance of the two LECs $C_{3S1}^{BB'}$ and $\hat C_{3S1}^{BB'}$ in $V^{BB'}_{\textrm{CT}}(^3P_0)$ does not
%  imply ``flavor mixing'' because their  combination here is not equivalent to $V^{BB'}_{\textrm{CT}}(^3S_1)$.}

\begin{align}
  & V^{YY'}_{\textrm{CT}}(^1S_0) = \xi_B \left[ (C_1^{YY'}+C_2^{YY'}-6C_3^{YY'}+3C_4^{YY'})(1+R_p^2R_{p'}^2) + (3C_2^{YY'}+6C_3^{YY'}+C_4^{YY'}+C_5^{YY'})(R_p^2+R_{p'}^2) \right] \nonumber\\
  & \phantom{V^{YY'}_{\textrm{CT}}(^1S_0)}\equiv \xi_B \left[ C_{1S0}^{YY'}(1+R_p^2R_{p'}^2) + \hat C_{1S0}^{YY'}(R_p^2+R_{p'}^2) \right], \label{ctp1}
\end{align}
\begin{align}
  & V^{YY'}_{\textrm{CT}}(^3S_1) = \xi_B \left[ \frac{1}{9}(C_1^{YY'}+C_2^{YY'}+2C_3^{YY'}-C_4^{YY'})(9+R_p^2R_{p'}^2) + \frac{1}{3}(C_2^{YY'}+2C_3^{YY'}-C_4^{YY'}-C_5^{YY'})(R_p^2+R_{p'}^2) \right] \nonumber\\
  & \phantom{V^{YY'}_{\textrm{CT}}(^3S_1)} \equiv \xi_B \left[ \frac{1}{9}C_{3S1}^{YY'}(9+R_p^2R_{p'}^2) + \frac{1}{3}\hat C_{3S1}^{YY'}(R_p^2+R_{p'}^2) \right], \label{ctp2}
\end{align}
\begin{align}
  & V^{YY'}_{\textrm{CT}}(^3P_1) = \xi_B \left[ -\frac{4}{3}(C_1^{YY'}-2C_2^{YY'}+4C_3^{YY'}+2C_4^{YY'}-C_5^{YY'})R_pR_{p'} \right]  \nonumber\\
  & \phantom{V^{YY'}_{\textrm{CT}}(^3P_1)} \equiv \xi_B \left[ -\frac{4}{3}C_{3P1}^{YY'}R_pR_{p'} \right], \label{ctp3}
\end{align}
\begin{align}
  & V^{YY'}_{\textrm{CT}}(^3P_0) = \xi_B \left[ -2(C_1^{YY'}-4C_2^{YY'}-4C_4^{YY'}+C_5^{YY'})R_pR_{p'} \right]\nonumber\\
  & \phantom{V^{YY'}_{\textrm{CT}}(^3P_0) } = \xi_B  \left[ -2(-C_{1S0}^{YY'} - \hat C_{1S0}^{YY'} + 2C_{3S1}^{YY'} - 2\hat C_{3S1}^{YY'} )R_pR_{p'} \right], \label{ctp4}
\end{align}
\begin{align}
  & V^{YY'}_{\textrm{CT}}(^1P_1) = \xi_B \left[ -\frac{2}{3}(C_1^{YY'}+C_5^{YY'})R_pR_{p'} \right] \nonumber\\
  & \phantom{V^{YY'}_{\textrm{CT}}(^1P_1)} = \xi_B \left[ -\frac{2}{3}(C_{3S1}^{YY'} - \hat C_{3S1}^{YY'})R_pR_{p'} \right], \label{ctp5}
\end{align}
\begin{align}
  & V^{YY'}_{\textrm{CT}}(^3S_1-{}^3D_1) = \xi_B \left[ \frac{2}{9}\sqrt{2}(C_1^{YY'}+C_2^{YY'}+2C_3^{YY'}-C_4^{YY'})R_p^2R_{p'}^2  + \frac{2}{3}\sqrt{2}(C_2^{YY'}+2C_3^{YY'}-C_4^{YY'}-C_5^{YY'})R_p^2 \right] \nonumber\\
  & \phantom{V^{YY'}_{\textrm{CT}}(^3D_1-{}^3S_1) } = \xi_B \left[ \frac{2}{9}\sqrt{2}C_{3S1}^{YY'}R_p^2R_{p'}^2 + \frac{2}{3}\sqrt{2}\hat C_{3S1}^{YY'}R_p^2 \right], \label{ctp6}
\end{align}
\begin{align}
  & V^{YY'}_{\textrm{CT}}(^3D_1-{}^3S_1) = \xi_B \left[ \frac{2}{9}\sqrt{2}(C_1^{YY'}+C_2^{YY'}+2C_3^{YY'}-C_4^{YY'})R_p^2R_{p'}^2  + \frac{2}{3}\sqrt{2}(C_2^{YY'}+2C_3^{YY'}-C_4^{YY'}-C_5^{YY'})R_{p'}^2 \right] \nonumber\\
  & \phantom{V^{YY'}_{\textrm{CT}}(^3D_1-{}^3S_1) } = \xi_B \left[ \frac{2}{9}\sqrt{2}C_{3S1}^{YY'}R_p^2R_{p'}^2 + \frac{2}{3}\sqrt{2}\hat C_{3S1}^{YY'}R_{p'}^2 \right], \label{ctp7}
\end{align}
\begin{align}
  & V^{YY'}_{\textrm{CT}}(^3D_1) = \xi_B \left[ \frac{8}{9}(C_1^{YY'}+C_2^{YY'}+2C_3^{YY'}-C_4^{YY'})R_p^2R_{p'}^2 \right] \nonumber\\
  & \phantom{V^{YY'}_{\textrm{CT}}(^3D_1)} = \xi_B \left[ \frac{8}{9} C_{3S1}^{YY'}R_p^2R_{p'}^2 \right], \label{ctp8}
\end{align}
where $\xi_B=N_p^2N_{p'}^2$, $R_p=|\mbox{\boldmath $p$}|/(E_p+M_B)$, $R_{p'}=|\mbox{\boldmath $p$}'|/(E_{p'}+M_B)$ and $M_B=1080$ MeV stands for the SU(3) average mass of the octet baryons in the chiral limit.\footnote{The baryon mass difference is treated as a higher order correction in chiral perturbation theory.}
\mbox{\boldmath $p$} and \mbox{\boldmath $p$}$'$ denote the initial and final momenta, respectively. Note that the second line of Eq.~(\ref{ctp4}) for $V^{YY'}_{\textrm{CT}}(^3P_0)$ is only valid for $NN$ interactions, because the structures of the Lagrangians for $^1S_0$ and $^3S_1$ partial waves are different in $\Lambda N-\Sigma N$ systems, as shown in Eqs.~(\ref{lala1}-\ref{lasi2}).
To recover the potentials in the HB approach we simply take $R_p=R_{p'}= 0$ and $\xi_B= 1$. The independent potentials respecting SU(3) symmetry are shown in Table \ref{CTtm}.
\begin{table}[h]
%\footnotesize
\renewcommand{\arraystretch}{1.2}
\centering
\caption{Independent contact terms and LECs of strangeness $S=-1$ $YN$ system.}
\label{CTtm}
\begin{tabular}{L{2cm}C{1.5cm}L{3cm}L{3cm}L{3cm}}
\hline
\hline
        Channel     & I & \multicolumn{3}{c}{$V$} \\
\cline{3-5}
                    &   & $^1S_0$       & $^3P_1$  & $^3S_1$  \\
\hline
 $\Lambda N \rightarrow \Lambda N$ & $\frac{1}{2}$ & $V^{\Lambda\Lambda}_{1S0}$ & $V^{\Lambda\Lambda}_{3P1}$ & $V^{\Lambda\Lambda}_{3S1}$\\
 $\Lambda N \rightarrow \Sigma N$  & $\frac{1}{2}$ & $3(V^{\Lambda\Lambda}_{1S0}-V^{\Sigma\Sigma}_{1S0})$ & $3(V^{\Lambda\Lambda}_{3P1}-V^{\Sigma\Sigma}_{3P1})$ & $V^{\Lambda\Sigma}_{3S1}$ \\
 $\Sigma N \rightarrow \Sigma N$   & $\frac{1}{2}$ & $9V^{\Lambda\Lambda}_{1S0}-8V^{\Sigma\Sigma}_{1S0}$ & $9V^{\Lambda\Lambda}_{3P1}-8V^{\Sigma\Sigma}_{3P1}$ & $V^{\Lambda\Lambda}_{3S1}$ \\
 $\Sigma N \rightarrow \Sigma N$   & $\frac{3}{2}$ & $V^{\Sigma\Sigma}_{1S0}$ & $V^{\Sigma\Sigma}_{3P1}$ & $V^{\Sigma\Sigma}_{3S1}$   \\
%\hline
%\multicolumn{5}{l}{Independent LECs: $C_{1S0}^{\Lambda\Lambda}$, $\hat C_{1S0}^{\Lambda\Lambda}$, $C_{1S0}^{\Sigma\Sigma}$, $\hat C_{1S0}^{\Sigma\Sigma}$, $C_{3S1}^{\Lambda\Lambda}$, $\hat C_{3S1}^{\Lambda\Lambda}$, $C_{3S1}^{\Sigma\Sigma}$, $\hat C_{3S1}^{\Sigma\Sigma}$, $C_{3S1}^{\Lambda\Sigma}$, $\hat C_{3S1}^{\Lambda\Sigma}$, $C_{3P1}^{\Lambda\Lambda}$, $C_{3P1}^{\Sigma\Sigma}$.} \\
\hline
\hline
\end{tabular}
\end{table}
The analytical form of the potentials, e.g., $V^{\Lambda\Lambda}_{1S0}$, $V^{\Lambda\Lambda}_{1P1}$, can be obtained from Eqs.~(\ref{ctp1}-\ref{ctp8}).
Finally we have 12 independent LECs: $C_{1S0}^{\Lambda\Lambda}$, $\hat C_{1S0}^{\Lambda\Lambda}$, $C_{1S0}^{\Sigma\Sigma}$, $\hat C_{1S0}^{\Sigma\Sigma}$, $C_{3S1}^{\Lambda\Lambda}$, $\hat C_{3S1}^{\Lambda\Lambda}$, $C_{3S1}^{\Sigma\Sigma}$, $\hat C_{3S1}^{\Sigma\Sigma}$, $C_{3S1}^{\Lambda\Sigma}$, $\hat C_{3S1}^{\Lambda\Sigma}$, $C_{3P1}^{\Lambda\Lambda}$, $C_{3P1}^{\Sigma\Sigma}$. The other three LECs only contribute to the strangeness $S=-2$ system.

\subsubsection{One-pseudoscalar-meson-exchange potentials}
\begin{figure}
  \centering
  % Requires \usepackage{graphicx}
  \includegraphics[width=0.15\textwidth]{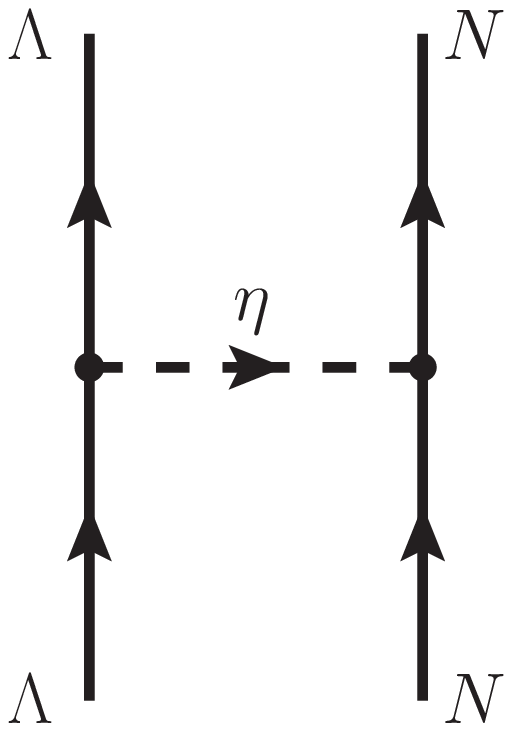}~~
  \includegraphics[width=0.15\textwidth]{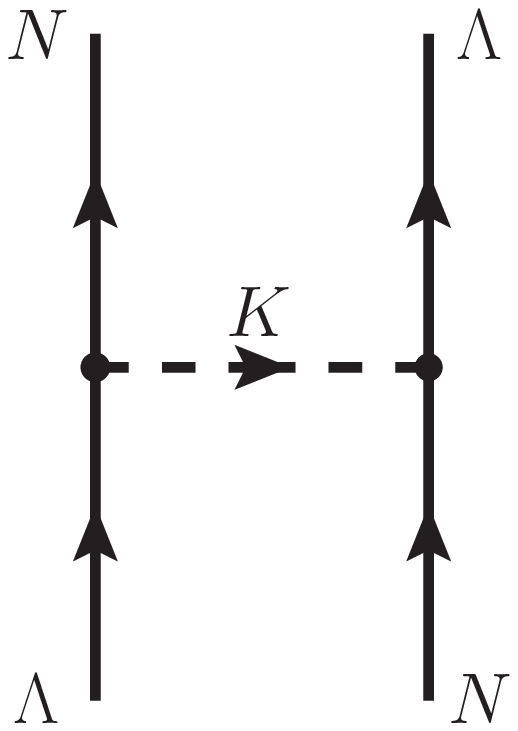}~~
  \includegraphics[width=0.15\textwidth]{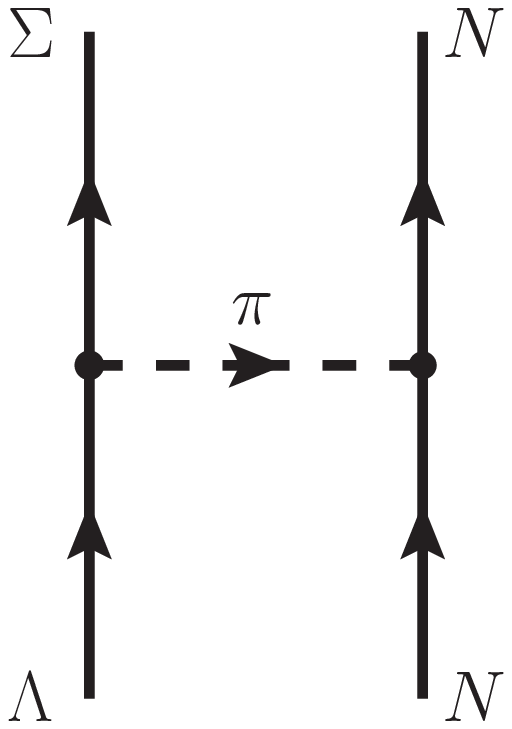}~~
  \includegraphics[width=0.15\textwidth]{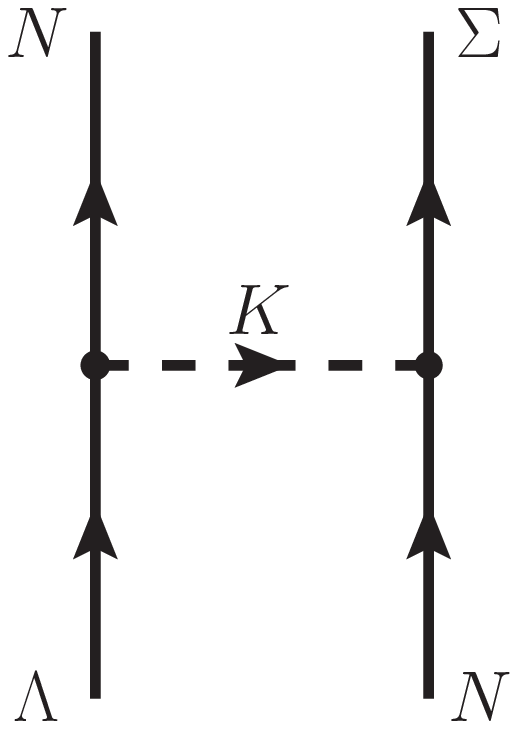}~~
  \includegraphics[width=0.15\textwidth]{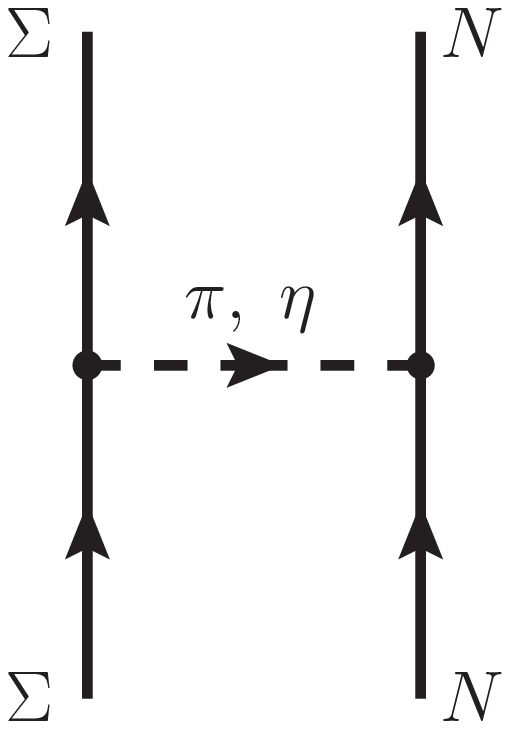}~~
  \includegraphics[width=0.15\textwidth]{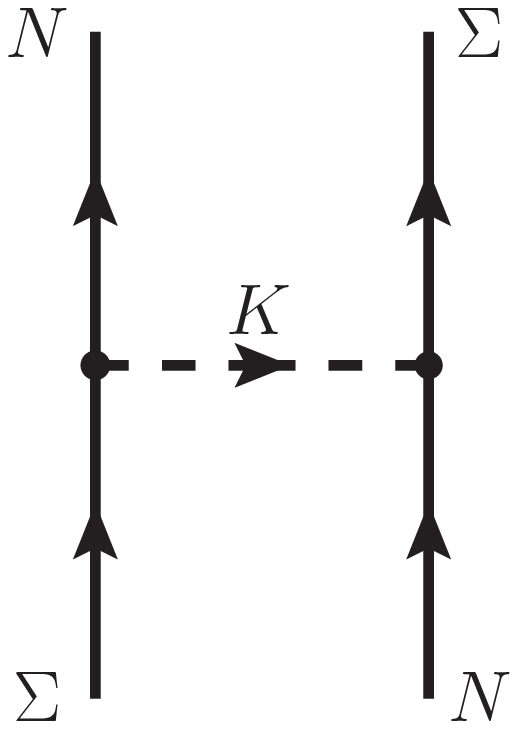}~~
  \caption{The one-pseudoscalar-meson exchange diagrams in the $\Lambda N-\Sigma N$ system.}\label{OME6}
\end{figure}

At LO, we have seven Feynman diagrams for strangeness $S=-1$ systems, as shown in Fig.~\ref{OME6}. The OPME potentials are derived from the covariant SU(3) meson-baryon Lagrangian,
\begin{align}\label{LMB1}
  &\mathcal{L}_{MB}^{(1)} =
  \mathrm{tr}\Bigg( \bar B \big(i\gamma_\mu D^\mu - M_B \big)B -\frac{D}{2} \bar B \gamma^\mu\gamma_5\{u_\mu,B\}
  -  \frac{F}{2}\bar{B} \gamma^\mu\gamma_5 [u_\mu,B]\Bigg)\, ,
\end{align}
where $D^\mu B = \partial_\mu B+[\Gamma_\mu,B] $ and $D$ and $F$ are the axial vector couplings. In the numerical analysis, we use
$D+F = g_A =1.277$ \cite{Patrignani:2016xqp} and $F/(F+D)=0.4$, where $g_A$ is the nucleon axial vector coupling constant. $\Gamma_\mu$  and $u_\mu$ are the vector and axial vector combinations of the pseudoscalar-meson fields and their derivatives,
\[
  \Gamma_\mu = \frac{1}{2}\left(u^\dag\partial_\mu u + u\partial_\mu u^\dag \right), \quad u_\mu=i(u^\dagger \partial_\mu u-u\partial_\mu u^\dagger)\, ,
\]
where $u^2= U = \exp\left(i\frac{\sqrt{2}\phi}{f_0}\right)$, with the pseudoscalar-meson decay constant $f_0\simeq f_\pi=92.2$ MeV \cite{Patrignani:2016xqp}, and the traceless matrix $\phi$ collecting the pseudoscalar-meson fields is:
\begin{equation}\label{MMatrix}
  \phi =
  \left(
   \begin{array}{ccc}
    \frac{\pi^0}{\sqrt{2}}+\frac{\eta}{\sqrt{6}} & \pi^+ & K^+ \\
    \pi^- & -\frac{\pi^0}{\sqrt{2}}+\frac{\eta}{\sqrt{6}} & K^0\\
    K^- & \bar K^0 & -\frac{2\eta}{\sqrt{6}}
  \end{array}
  \right).
\end{equation}
The potentials for OPME can be expressed in a generic form:
\begin{equation}\label{VOMEP}
V_\mathrm{OPME} = -N_{B_1B_3\phi}N_{B_2B_4\phi}
   \frac{(\bar u_3 \gamma^\mu \gamma_5 q_\mu u_1) (\bar u_4 \gamma^\nu \gamma_5 q_\nu u_2)}
        {q^2-m^2}\mathcal{I}_{B_1B_2\rightarrow B_3B_4}\, ,
\end{equation}
where $q=p'-p$ is the momentum transfer, $q^2=(E_{p'}-E_p)^2-(\mbox{\boldmath $p$}'-\mbox{\boldmath $p$})^2$, and $m$ is the mass of the exchanged pseudoscalar meson. The SU(3) coefficient $N_{BB'\phi}$ and isospin factor $\mathcal{I}_{B_1B_2\rightarrow B_3B_4}$ are listed in Refs.~\cite{Polinder:2006zh,Li:2016paq}. The retardation effects are included in the denominator. Just like the contact terms, Eqs.~(\ref{ctsym}-\ref{ctp8}), the average baryon mass $M_B=1080$ MeV is used in the baryon spinors $u(\bar u)$ and energies $E_{p(p')}$.
One can easily obtain $V_\mathrm{OPME}$ in the $\left|LSJ\right>$ basis following the same procedure as that for the contact terms. We note that by the mass differences of the exchanged mesons\footnote{We have used $m_\pi=138.039$ MeV, $m_K=495.645$ MeV and $m_\eta=547.853$ MeV in the numerical calculations.} the SU(3) symmetry is broken.

In our covariant power counting scheme we keep the complete form of the Dirac spinors and do not perform expansions in terms of small external three momenta, different from what done in the HB approach. In relativistic atomic and nuclear structure studies, the small components of the Dirac spinors have been shown to play an important role, mostly of a dynamical nature. As we will see
below, they also play an important role in the present study and result in a good description of $YN$ scattering data. Because the small components are retained, once written in terms of three-momenta and Pauli matrices, the relativistic potential contains terms of higher chiral order in the HB language, similar to the one-baryon sector in covariant chiral perturbation theory.
Furthermore, we can see that the LO potentials obtained in the EG approach are
the same as those of the HB approach~\cite{Li:2016paq}, different from the relativistic potentials.

\subsection{Scattering equation}
The infrared enhancement in two-baryon propagations gives the theoretical argument for low-energy baryon-baryon interactions to be non-perturbative~\cite{Weinberg:1991um}. As a result, one needs to iterate the potentials in the Bethe-Salpeter equation. In practice this is difficult. A three-dimensional reduction of the Bethe-Salpeter equation is often used~\cite{Woloshyn:1974wm}. In the present work, following Ref.~\cite{Li:2016paq}, we use the coupled-channel Kadyshevsky equation
%\footnote{Since in the present case, we set the zero component of $p''$ at zero in the potentials in deriving Eq.~(\ref{SEK}) from the Bethe-Salpeter equation. Therefore,
%it should be more properly referred to as the Thompson I equation~\cite{Thompson:1970wt}.} :
%\end{multicols}
\begin{align}\label{SEK}
  & T_{\rho\rho'}^{\nu\nu',J}(\mbox{\boldmath $p$}',\mbox{\boldmath $p$};\sqrt{s})
  =
   V_{\rho\rho'}^{\nu\nu',J}(\mbox{\boldmath $p$}',\mbox{\boldmath $p$})
   %\notag\\
   % &\qquad
   +
  \sum_{\rho'',\nu''}\int_0^\infty \frac{dp''p''^2}{(2\pi)^3} \frac{M_{B_{1,\nu''}}M_{B_{2,\nu''}}~ V_{\rho\rho''}^{\nu\nu'',J}(\mbox{\boldmath $p$}',\mbox{\boldmath $p$}'')~
   T_{\rho''\rho'}^{\nu''\nu',J}(\mbox{\boldmath $p$}'',\mbox{\boldmath $p$};\sqrt{s})}{E_{1,\nu''}E_{2,\nu''}
  \left(\sqrt{s}-E_{1,\nu''}-E_{2,\nu''}+i\epsilon\right)},
\end{align}
%\begin{multicols}{2}
where $\sqrt{s}$ is the total energy of the baryon-baryon system in the center-of-mass frame
%,
%$\mbox{\boldmath $q$}_{\nu''}$ is the relativistic on-shell momentum defined by $\sqrt{s} = \sqrt{\mbox{\boldmath $q$}^2_{\nu''}+M_{B_{1,\nu''}}^2} + \sqrt{\mbox{\boldmath $q$}^2_{\nu''}+M_{B_{2,\nu''}}^2}$, with $B_{1,\nu''}$ and $B_{2,\nu''}$  intermediate state baryons,
and $E_{n,\nu''}=\sqrt{\mbox{\boldmath $p$}''+M_{B_{n,\nu''}}}$, $(n=1,2)$. The labels $\nu,\nu',\nu''$ denote the particle channels, and $\rho,\rho',\rho''$ denote the partial waves.
Relativistic kinematics is used throughout to relate the laboratory momenta to the center-of-mass momenta.

To regularize the integration in the high-momentum region, baryon-baryon potentials are multiplied with an exponential form factor,
\begin{equation}\label{EF}
  f_{\Lambda_F}(\mbox{\boldmath $p$},\mbox{\boldmath $p$}') = \exp \left[-\left(\frac{\mbox{\boldmath $p$}}{\Lambda_F}\right)^{2n}-\left(\frac{\mbox{\boldmath $p$}'}{\Lambda_F}\right)^{2n}\right] \, ,
\end{equation}
where $n=2$~\cite{Epelbaum:2004fk}.
Note that Eq.~(\ref{EF}) is not a covariant cutoff function. Although there exist covariant cutoff functions of $q^2$, they are not favored in constructing chiral forces because they will introduce additional angular dependence to partial wave potentials and thus affect the interpretation of contact interactions. It would be interesting to construct a separable and covariant cutoff function and study its consequences in the future.

The Kadyshevsky equation is solved in the particle basis in order to properly account for the physical thresholds and the Coulomb force in charged channels. The latter is treated with the Vincent-Phatak method~\cite{Vincent:1974zz}.

\section{Fitting procedure\label{sec_fitting}}

In our approach, there are 12 LECs that need to be pinned down by fitting to the 36 $YN$ scattering data points as done  in Ref.~\cite{Li:2016paq}, which consist of 35 cross sections~\cite{SechiZorn:1969hk,Alexander:1969cx,Eisele:1971mk,Engelmann:1966} and a $\Sigma^-p$ inelastic capture ratio at rest~\cite{Hepp:1968zza}.

Due to the poor quality of experimental data,
it is customary to consider the hypertriton $^3_\Lambda$H binding energy \cite{Juric:1973zq,Davis:1991zpu} as a further constraint, which is crucial in fixing the relative strength of the $^1S_0$ and $^3S_1$ contributions to $\Lambda p$ scattering.
However, we are unable to perform a 3-body calculation at present, so we use as benchmarks the $\Lambda p$ $S$-wave scattering lengths extracted in the LO~\cite{Polinder:2006zh} and next-to-leading order (NLO)~\cite{Haidenbauer:2013oca} HB calculations, mainly because they combine to describe the hypertriton very well~\cite{Nogga:2013pwa}.
%\textbf{The $\Lambda p$ $^3S_1$ contribution can be relatively well determined by only the scattering data while the $^1S_0$ part is mainly decided by the hypertriton binding energy} \cite{Korpa:2001au}.
In addition, it seems necessary that $a^{\Lambda p}_{1S0}$ should be neither smaller nor too much larger than $a^{\Lambda p}_{3S1}$, as shown in Ref.~\cite{Tominaga:1998iy}.

 %However, we are unable to perform a 3-body calculation at present, therefore we use the $\Lambda p$ $S$-wave scattering lengths as further constraints. At present, their values have not been precisely determined (see, e.g., Ref.~\cite{Hauenstein:2016zys} for a latest study). As a result, we take the scattering lengths from the  LO \cite{Polinder:2006zh} and next-to-leading order (NLO)~\cite{Haidenbauer:2013oca} Weinberg approach  as references, because both of them describe very well the hypertriton \cite{Nogga:2013pwa}.

Another constraint that should be considered is the $\Sigma^+p$ $^3S_1$ scattering length. A repulsive $\Sigma N$ interaction with isospin $I=3/2$ is obtained from recent experiments~\cite{Batty:1994yx,Mares:1995bm,Bart:1999uh,Noumi:2001tx,Saha:2004ha,Kohno:2006iq,Dabrowski:2008zza}.  In addition, the conventional $G$-matrix calculations \cite{Haidenbauer:2014uua} indicate that the $^3S_1$ partial wave for $I=3/2$ $\Sigma N$ should be at least moderately repulsive, therefore in our fits we require a positive $a^{\Sigma^+ p}_{3S1}$.

Previous works in $\chi$EFT~\cite{Polinder:2006zh,Haidenbauer:2013oca,Li:2016paq} showed that the optimum cutoff $\Lambda_F$ may be around
600 MeV. Therefore we first tentatively fix $\Lambda_F$ at $600$ MeV. With this cutoff  we find that
the best description of the experimental data yields $a^{\Lambda p}_{3S1}\approx-1.30\pm0.02$ fm  and $a^{\Lambda p}_{1S0}\approx-2.44^{+0.16}_{-0.54}$ fm. These numbers are between the LO and NLO HB results, which are $a^{\Lambda p}_{3S1}=-1.23$ fm (LO), $a^{\Lambda p}_{3S1}=-1.54$ fm (NLO),  $a^{\Lambda p}_{1S0}=-1.91$ fm (LO), and $a^{\Lambda P}_{1S0}=-2.91$ fm (NLO). Best fits within $\Lambda_F=500-850$ MeV yield similar scattering lengths. In the results presented below, we fix $a^{\Lambda p}_{3S1}=-1.32$ fm and $a^{\Lambda p}_{1S0}=-2.44$ fm.\footnote{We have chosen a larger $a^{\Lambda p}_{3S1}$ given the fact that most phenomenological studies seem to prefer a larger scattering length in this channel.} It should be noted that at present we could in principle choose other  combinations within the uncertainties allowed in the best fits. To fix them uniquely, more experimental inputs are needed .

We have made an attempt at a combined fit to the $NN$ and $YN$ data, in which strict SU(3) symmetry was imposed upon the contact terms so that no additional LECs are needed. However, we failed to describe the $NN$ and $YN$ data  simultaneously. As a result, consistent with previous NLO results in the HB approach~\cite{Haidenbauer:2013oca}, we conclude that one needs to treat  SU(3) symmetry breaking more carefully in order to simultaneously describe both the $NN$ and the $YN$ systems in $\chi$EFT.

\section{Results and discussion\label{sec_results}}

\begin{figure}
  \centering
  % Requires \usepackage{graphicx}
  \includegraphics[width=0.5\textwidth]{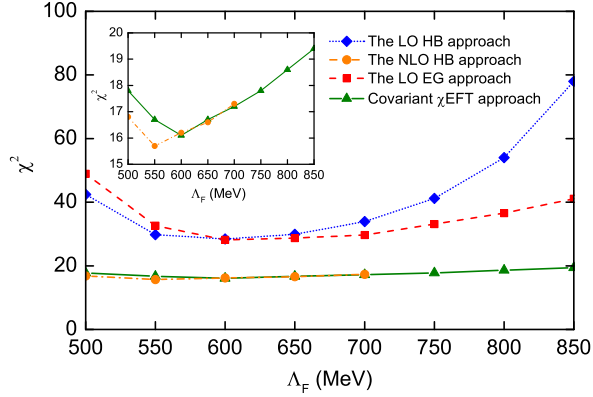}
  \caption{$\chi^2$ as a function of the cutoff in the covariant $\chi$EFT approach at LO (green solid line), the HB approach at LO (blue dotted line), NLO (orange dashed-dotted line)~\cite{Haidenbauer:2013oca} and the EG approach at LO (red dashed line).}\label{chi2}
\end{figure}

With the three additional constraints $a^{\Lambda p}_{1S0}=-2.44$ fm, $a^{\Lambda p}_{3S1}=-1.32$ fm and $a^{\Sigma^+ p}_{3S1} > 0$ as explained above, we
perform a fit to the 36 scattering data points while varying the cutoff $\Lambda_F$. The dependence of $\chi^2$ on the cutoff is shown in Fig.~\ref{chi2}, in comparison with other approaches. One can see that our new covariant $\chi$EFT approach shows a clear improvement in describing the $YN$ data compared with the HB and EG approach at LO, and the cutoff dependence is much mitigated, both of which are comparable with the NLO HB approach~\cite{Haidenbauer:2013oca}.
%, which, however, has 23 LECs.
The minimum value of the $\chi^2$ is about 16.1, located at $\Lambda_F=550-650$ MeV.
Note that the NSC97a-f \cite{Rijken:1998yy} models, which provide the best description among the phenomenological potentials of the 36 scattering data points,  also have a $\chi^2$ around 16.

The best fitted LECs obtained with $\Lambda_F=600$ MeV are listed in Table \ref{tab_LECs600}. Since the LECs in the $\Lambda p$ $^1S_0$ partial wave cannot  be uniquely determined, as mentioned previously, we only show a typical case here. One should note that these LECs are certain combinations of those appearing in the Lagrangians, and hence they are not necessarily of the same order of magnitude (see, e.g., Refs.~\cite{Polinder:2006zh,Haidenbauer:2013oca}).

\begin{figure}
  \centering
  % Requires \usepackage{graphicx}
  \includegraphics[width=1\textwidth]{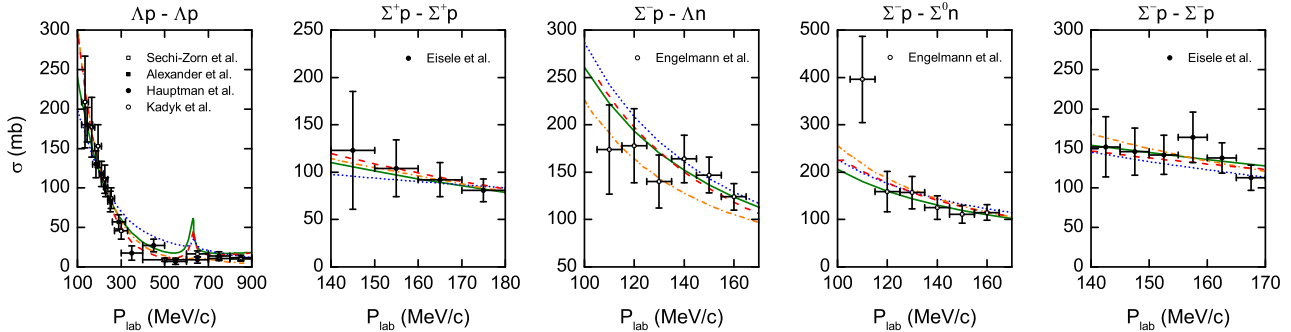}
  \caption{Cross sections in the covariant $\chi$EFT approach (green solid lines) and HB approach (blue dotted lines) at LO as functions of the laboratory momentum at $\Lambda_F=600$ MeV. For reference, the NSC97f \cite{Rijken:1998yy} (red dashed lines) and J\"ulich 04~\cite{Haidenbauer:2005zh} (orange dashed-dotted lines) results are also shown. The experimental data are taken from Sechi-Zorn \textit{et al}.~\cite{SechiZorn:1969hk}, Alexander \textit{et al}.~\cite{Alexander:1969cx},
  Eisele \textit{et al}.~\cite{Eisele:1971mk}, Engelmann \textit{et al}.~\cite{Engelmann:1966}, Hauptman \textit{et al}.~\cite{Hauptman:1977hr} and Kadyk \textit{et al}.~\cite{Kadyk:1971tc}.}\label{cs}
\end{figure}

In Fig.~\ref{cs} we compare the descriptions of the experimental cross sections that we have used in the fitting procedure with the LO HB approach. The NSC97f \cite{Rijken:1998yy} and J\"ulich 04 results~\cite{Haidenbauer:2005zh} are also shown for comparison. It is clear that the covariant $\chi$EFT approach can reproduce the experimental data rather well. The cusp at the $\Sigma N$ threshold in the $\Lambda p \rightarrow \Lambda p$ reaction is also  reproduced well. Note that the experimental data with $P_{\textrm{lab}}>300$ MeV are not used in the fitting procedure.

\begin{figure}
  \centering
  % Requires \usepackage{graphicx}
  \includegraphics[width=1.0\textwidth]{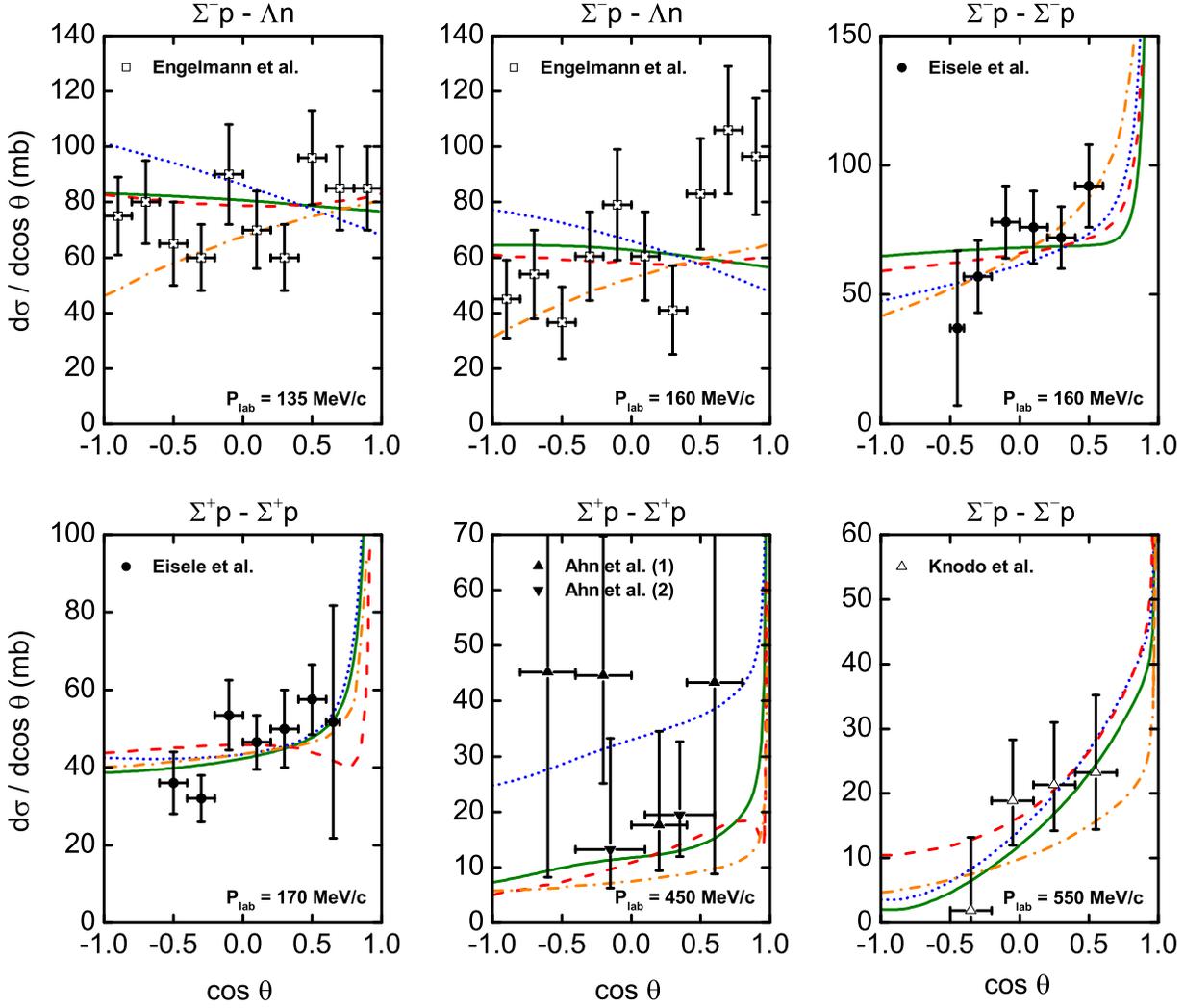}
  \caption{Differential cross sections as a function of cos$\theta$ at various laboratory momenta $P_{\textrm{lab}}$, where $\theta$ is the center-of-mass scattering angle. The covariant $\chi$EFT approach is shown by the green solid lines, the HB approach at LO by the blue dotted lines,  the NSC97f \cite{Rijken:1998yy} results by the red dashed lines, and the J\"ulich 04~\cite{Haidenbauer:2005zh} results by the orange dashed-dotted lines. The experimental data are taken from
  Engelmann \textit{et al.}~\cite{Engelmann:1966}, Eisele \textit{et al.}~\cite{Eisele:1971mk}, Ahn \textit{et al.}~\cite{Ahn:2005gb,Ahn:1997wa} and Kohno \textit{et al.}~\cite{Kohno:1999nz}.}\label{dcs}
\end{figure}

\begin{table*}
\centering
 \caption{Low-energy constants (in units of $10^4$ GeV$^{-2}$) at $\Lambda_F=600$ MeV in the covariant $\chi$EFT approach.}
\footnotesize
\renewcommand\arraystretch{2}
   \begin{tabular}{ccccccccccccc}
  \hline
  \hline
  % after \\: \hline or \cline{col1-col2} \cline{col3-col4} ...
  LECs & $C^{\Lambda \Lambda}_{1S0}$ & $C^{\Sigma \Sigma}_{1S0}$ & $C^{\Lambda \Lambda}_{3S1}$ & $C^{\Sigma \Sigma}_{3S1}$ & $C^{\Lambda \Sigma}_{3S1}$ & $\hat C^{\Lambda \Lambda}_{1S0}$ & $\hat C^{\Sigma \Sigma}_{1S0}$ & $\hat C^{\Lambda \Lambda}_{3S1}$ & $\hat C^{\Sigma \Sigma}_{3S1}$ & $\hat C^{\Lambda \Sigma}_{3S1}$ & $C^{\Lambda \Lambda}_{3P1}$ & $C^{\Sigma \Sigma}_{3P1}$ \\
  \hline
   & $-0.0096$ & $-0.0276$ & $0.0110$ & $0.0872$ & $0.0257$ & $4.2463$ & $4.6182$ & $0.3660$ & $-0.4132$ & $0.8499$ & $0.2044$ & $0.2616$\\
  \hline
  \hline
\end{tabular}\label{tab_LECs600}
 \end{table*}
Due to the lack of near-threshold experimental data, the value of the $\Lambda p \rightarrow \Lambda p$ cross section at rest is not yet known. Our value is about 350 mb, which is smaller than the two phenomenological models.
%The $\Lambda N-\Sigma N$ coupling is always crucial in hypernuclear structure calculations~\cite{Hiyama:2014cua}, the property of which can be reflected partially from the $\Sigma^-p \rightarrow \Lambda n$ reaction.
Our result in the $\Sigma^-p \rightarrow \Lambda n$ reaction is similar to the LO HB approach and NSC97f results, but quite different from the J\"ulich 04 model. This channel can partially reflect the nature of $\Lambda N-\Sigma N$ coupling, which is crucial in hypernuclear structure calculations~\cite{Hiyama:2014cua}. It is interesting to note that the J\"ulich 04 model predicts an overbound $\Lambda$ single particle potential $U_\Lambda(0)$ in $G$-matrix calculations. On the other hand, the results from the former two are much closer to the empirical value, c.f. Ref.~\cite{Haidenbauer:2014uua} and references therein. In addition, the differential cross sections shown in Fig.~\ref{dcs} are also well predicted within experimental uncertainties, although those data are not taken into account in the fitting procedure.

\begin{figure}
  \centering
  % Requires \usepackage{graphicx}
  \includegraphics[width=1\textwidth]{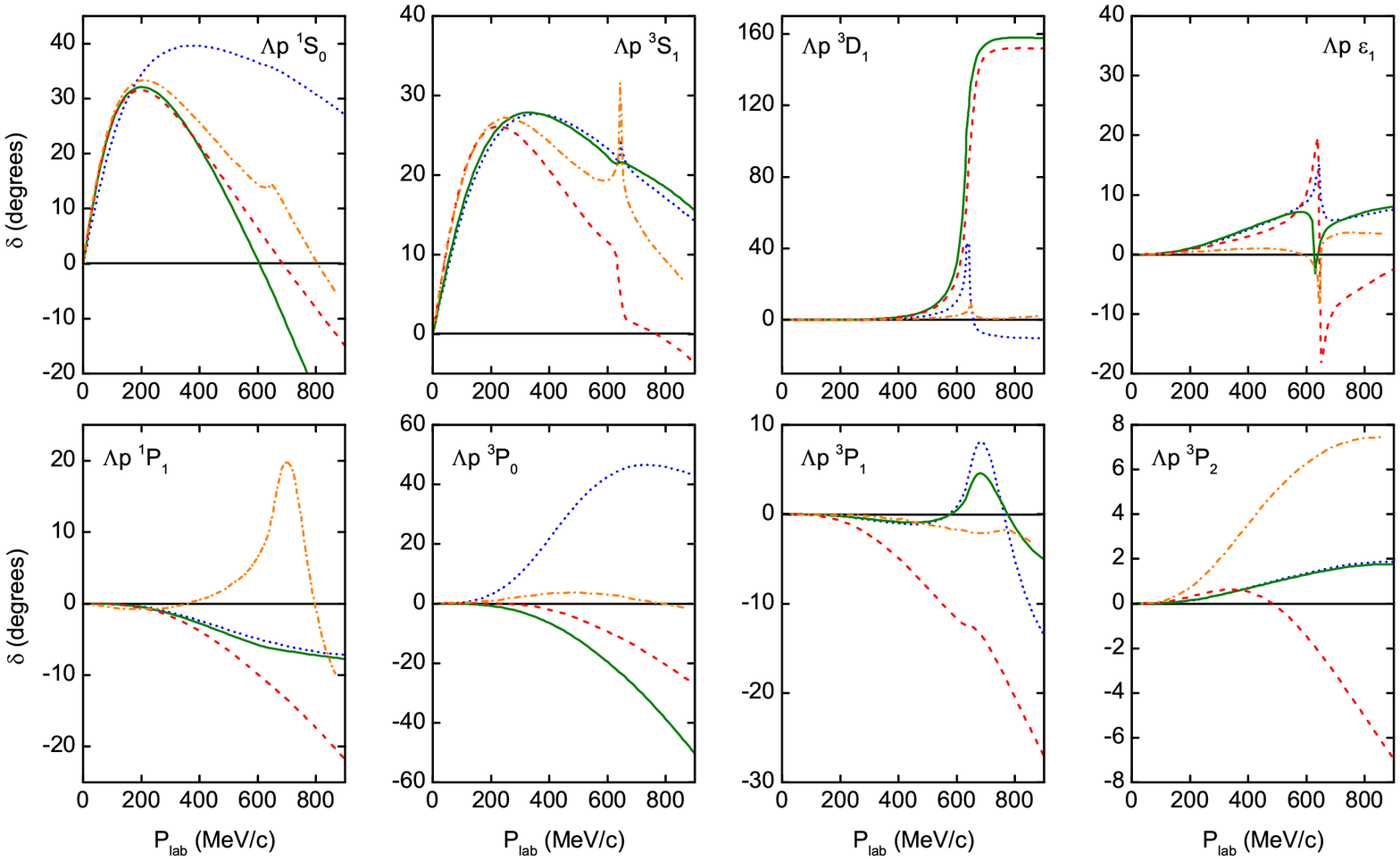}
  \caption{$\Lambda p$ $S$- and $P$-wave phase shifts in the covariant $\chi$EFT approach (green solid lines) and HB approach (blue dotted lines) at LO as functions of the laboratory momentum at $\Lambda_F=600$ MeV. For reference, the NSC97f \cite{Rijken:1998yy} (red dashed lines) and J\"ulich 04~\cite{Haidenbauer:2005zh} (orange dashed-dotted lines) results are also shown.}\label{LaPps}
\end{figure}

\begin{figure}
  \centering
  % Requires \usepackage{graphicx}
  \includegraphics[width=1\textwidth]{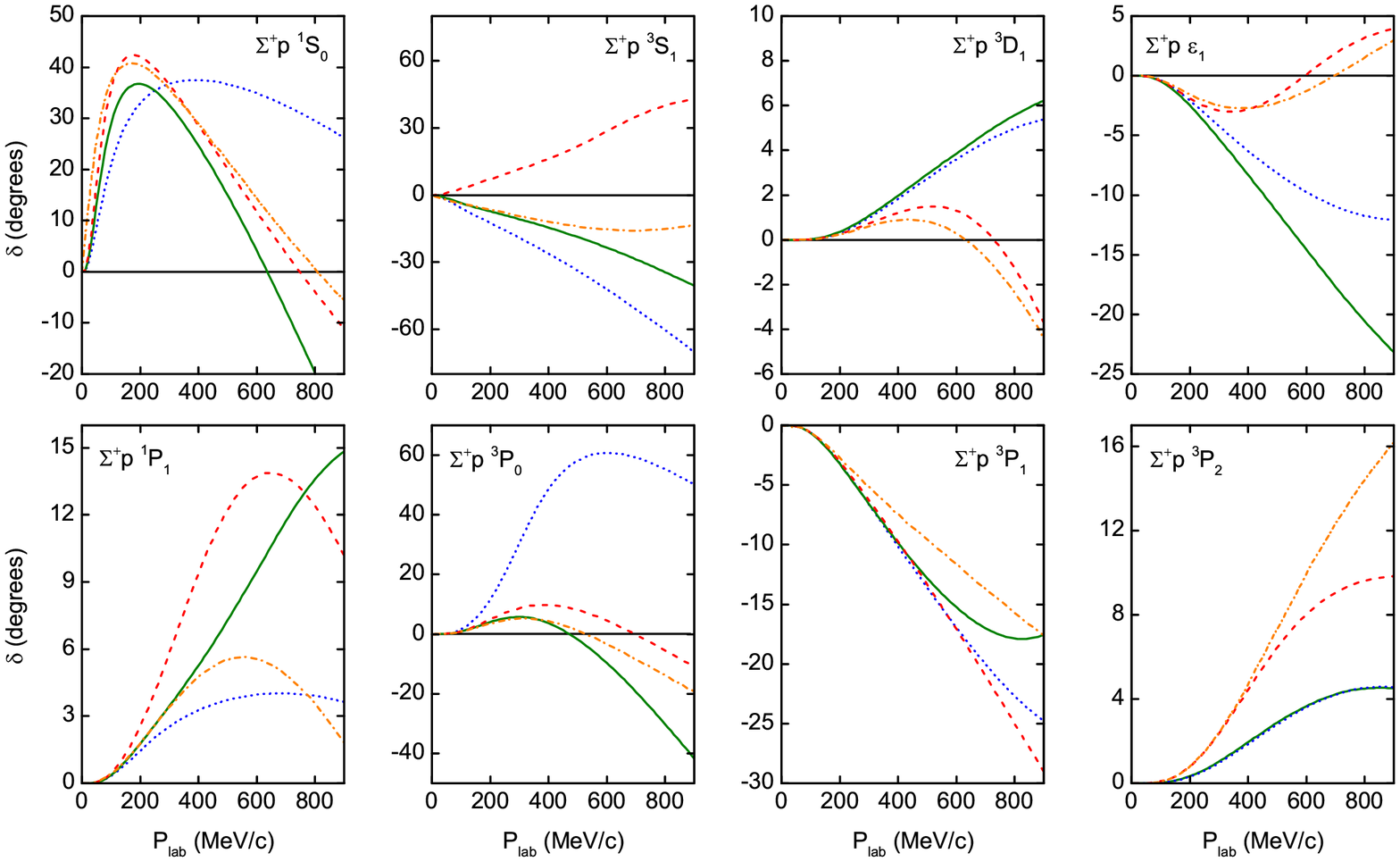}
  \caption{$\Sigma^+p$ $S$- and $P$-wave phase shifts in the covariant $\chi$EFT approach (green solid lines) and HB approach (blue dotted lines) at LO as functions of the laboratory momentum at $\Lambda_F=600$ MeV. For reference, the NSC97f \cite{Rijken:1998yy} (red dashed lines) and J\"ulich 04~\cite{Haidenbauer:2005zh} (orange dashed-dotted lines) results are also shown.}\label{SiPps}
\end{figure}

$S$- and $P$-wave phase shifts of $\Lambda p$ and $\Sigma^+ p$ reactions are shown in Figs.~\ref{LaPps}-\ref{SiPps}. The $^1S_0$ and $^3P_0$ phase shifts are quite different from those of the LO HB approach, but the $^3P_2$ phase shifts are similar, where only OPME terms contribute. Furthermore, the $^1S_0$ phase shifts are similar to those of the NLO HB approach~\cite{Haidenbauer:2013oca}.

The improved description of the scattering data by the covariant $\chi$EFT scheme for the most part arises from the contact terms. In the LO HB approach, contact terms only appear in central and spin-spin potentials without momentum dependence, which only contribute to the $^1S_0$ and $^3S_1$ partial waves. In covariant power counting, tensor, spin-orbit and quadratic spin-orbit terms appear at LO in addition to the central and spin-spin terms, namely the momentum dependent terms of $R_{p~(p')}^2$ in Eqs.~(\ref{ctp1}-\ref{ctp8}). These terms are responsible for the improved description. On the other hand, relativistic corrections to the OPME terms are small. As a result, phase shifts of higher partial waves where only such terms contribute are similar in the covariant and HB approaches at LO. A related discussion for the $NN$ sector can be found in Ref.~\cite{Ren:2016jna}.


\section{Summary and outlook\label{sec_summary}}

We have studied strangeness $S = -1$ hyperon-nucleon scattering at leading order in a covariant framework of chiral effective field theory. Starting from the covariant chiral Lagrangian, the small components of the baryon spinors are retained in deriving the potentials in order to preserve Lorentz invariance. Strict SU(3) symmetry is imposed on the contact terms, which yield 12 independent low energy constants. SU(3) symmetry is broken in the one-pseudoscalar-meson-exchange potentials because of the mass difference of exchanged mesons. The potentials are iterated using the Kadyshevsky equation. A quite satisfactory description of the 36 hyperon-nucleon scattering data points is obtained and the cutoff dependence is shown to be mitigated, both of which are comparable with the next-to-leading order heavy baryon approach. However, one cannot achieve a simultaneous description of the nucleon-nucleon phase shifts and strangeness $S=-1$ hyperon-nucleon scattering data at leading order. The relativistic interactions obtained in the this work may provide essential inputs to relativistic hypernuclear structure studies, e.g., relativistic Brueckner-Hartree-Fork theory in many-body systems.

%\end{multicols}

\vspace{10mm}

\vspace{-1mm}
\centerline{\rule{80mm}{0.1pt}}
\vspace{2mm}

%\begin{multicols}{2}

%\end{multicols}

\clearpage

\end{CJK*}

\begin{thebibliography}{500}

%\cite{GellMann:1953zza}
\bibitem{GellMann:1953zza}
  M.~Gell-Mann,
  %``Isotopic Spin and New Unstable Particles,''
  Phys.\ Rev.\  {\bf 92}, 833 (1953).
  %doi:10.1103/PhysRev.92.833
  %%CITATION = doi:10.1103/PhysRev.92.833;%%
  %234 citations counted in INSPIRE as of 12 Dec 2016

%\cite{Nakano:1953zz}
\bibitem{Nakano:1953zz}
  T.~Nakano and K.~Nishijima,
  %``Charge Independence for V-particles,''
  Prog.\ Theor.\ Phys.\  {\bf 10}, 581 (1953).
  %doi:10.1143/PTP.10.581
  %%CITATION = doi:10.1143/PTP.10.581;%%
  %202 citations counted in INSPIRE as of 25 Oct 2016

%\cite{Danysz:1953xy}
\bibitem{Danysz:1953xy}
  M.~Danysz, and J.~Pniewski,
  %``Delayed disintegration of a heavy nuclear fragment: I,''
  Philos.\ Mag.\ Ser.\ 5 {\bf 44}, 348 (1953).
  %dx.doi.org/10.1080/14786440308520318
  %%CITATION = dx.doi.org/10.1080/14786440308520318;%%
  %??? citations counted in INSPIRE as of 25 Oct 2016

%\cite{Bedjidian:1979qh}
\bibitem{Bedjidian:1979qh}
  M.~Bedjidian {\it et al.} [CERN-Lyon-Warsaw Collaboration],
  %``Further Investigation of the $\gamma$ Transitions in $^{4}$H ($\Lambda$) and $^{4}$He ($\Lambda$) Hypernuclei,''
  Phys.\ Lett.\  {\bf 83B}, 252 (1979).
  %doi:10.1016/0370-2693(79)90697-X
  %%CITATION = doi:10.1016/0370-2693(79)90697-X;%%
  %75 citations counted in INSPIRE as of 26 Oct 2016

%\cite{Jaffe:1976yi}
\bibitem{Jaffe:1976yi}
  R.~L.~Jaffe,
  %``Perhaps a Stable Dihyperon,''
  Phys.\ Rev.\ Lett.\  {\bf 38}, 195 (1977)
  Erratum: [Phys.\ Rev.\ Lett.\  {\bf 38}, 617 (1977)].
  %doi:10.1103/PhysRevLett.38.195
  %%CITATION = doi:10.1103/PhysRevLett.38.195;%%
  %1123 citations counted in INSPIRE as of 26 Oct 2016

%\cite{Nogga:2001ef}
\bibitem{Nogga:2001ef}
  A.~Nogga, H.~Kamada and W.~Gl\"ockle,
  %``The Hypernuclei (Lambda) He-4 and (Lambda) He-4: Challenges for modern hyperon nucleon forces,''
  Phys.\ Rev.\ Lett.\  {\bf 88}, 172501 (2002)
  %doi:10.1103/PhysRevLett.88.172501
  [nucl-th/0112060].
  %%CITATION = doi:10.1103/PhysRevLett.88.172501;%%
  %89 citations counted in INSPIRE as of 26 Oct 2016

%\cite{Gazda:2015qyt}
\bibitem{Gazda:2015qyt}
  D.~Gazda and A.~Gal,
  %``Ab initio Calculations of Charge Symmetry Breaking in the $A=4$ Hypernuclei,''
  Phys.\ Rev.\ Lett.\  {\bf 116}, 122501 (2016)
  %doi:10.1103/PhysRevLett.116.122501
  [arXiv:1512.01049 [nucl-th]].
  %%CITATION = doi:10.1103/PhysRevLett.116.122501;%%
  %13 citations counted in INSPIRE as of 26 Oct 2016

%\cite{Beane:2010hg}
\bibitem{Beane:2010hg}
  S.~R.~Beane {\it et al.} [NPLQCD Collaboration],
  %``Evidence for a Bound H-dibaryon from Lattice QCD,''
  Phys.\ Rev.\ Lett.\  {\bf 106}, 162001 (2011)
  %doi:10.1103/PhysRevLett.106.162001
  [arXiv:1012.3812 [hep-lat]].
  %%CITATION = doi:10.1103/PhysRevLett.106.162001;%%
  %192 citations counted in INSPIRE as of 26 Oct 2016

%\cite{Inoue:2010es}
\bibitem{Inoue:2010es}
  T.~Inoue {\it et al.} [HAL QCD Collaboration],
  %``Bound H-dibaryon in Flavor SU(3) Limit of Lattice QCD,''
  Phys.\ Rev.\ Lett.\  {\bf 106}, 162002 (2011)
  %doi:10.1103/PhysRevLett.106.162002
  [arXiv:1012.5928 [hep-lat]].
  %%CITATION = doi:10.1103/PhysRevLett.106.162002;%%
  %217 citations counted in INSPIRE as of 26 Oct 2016

%\cite{Haidenbauer:2011ah}
\bibitem{Haidenbauer:2011ah}
  J.~Haidenbauer and U.~-G.~Mei\ss ner,
  %``To bind or not to bind: The H-dibaryon in light of chiral effective field theory,''
  Phys.\ Lett.\ B {\bf 706}, 100 (2011)
  %doi:10.1016/j.physletb.2011.10.070
  [arXiv:1109.3590 [hep-ph]].
  %%CITATION = doi:10.1016/j.physletb.2011.10.070;%%
  %34 citations counted in INSPIRE as of 26 Oct 2016

%\cite{Yamaguchi:2016kxa}
\bibitem{Yamaguchi:2016kxa} 
  Y.~Yamaguchi and T.~Hyodo,
  %``Quark-mass dependence of the H dibaryon in $\Lambda\Lambda$ scattering,''
  Phys.\ Rev.\ C {\bf 94}, 065207 (2016)
  %doi:10.1103/PhysRevC.94.065207
  [arXiv:1607.04053 [hep-ph]].
  %%CITATION = doi:10.1103/PhysRevC.94.065207;%%
  %10 citations counted in INSPIRE as of 04 Jan 2018


%\cite{Gogami:2015tvu}
\bibitem{Gogami:2015tvu}
  T.~Gogami {\it et al.},
  %``High resolution spectroscopic study of $^{10}_{\Lambda}$Be,''
  Phys.\ Rev.\ C {\bf 93}, 034314 (2016)
  %doi:10.1103/PhysRevC.93.034314
  [arXiv:1511.04801 [nucl-ex]].
  %%CITATION = doi:10.1103/PhysRevC.93.034314;%%
  %9 citations counted in INSPIRE as of 28 Feb 2017

%\cite{Esser:2015trs}
\bibitem{Esser:2015trs}
  A.~Esser {\it et al.} [A1 Collaboration],
  %``Observation of $_\Lambda ^4$H Hyperhydrogen by Decay-Pion Spectroscopy in Electron Scattering,''
  Phys.\ Rev.\ Lett.\  {\bf 114}, 232501 (2015)
  %doi:10.1103/PhysRevLett.114.232501
  [arXiv:1501.06823 [nucl-ex]].
  %%CITATION = doi:10.1103/PhysRevLett.114.232501;%%
  %28 citations counted in INSPIRE as of 28 Feb 2017

%\cite{Yamamoto:2015avw}
\bibitem{Yamamoto:2015avw}
  T.~O.~Yamamoto {\it et al.} [J-PARC E13 Collaboration],
  %``Observation of Spin-Dependent Charge Symmetry Breaking in $\Lambda N$ Interaction: Gamma-Ray Spectroscopy of $^4_{\Lambda }$He,''
  Phys.\ Rev.\ Lett.\  {\bf 115}, 222501 (2015)
  %doi:10.1103/PhysRevLett.115.222501
  [arXiv:1508.00376 [nucl-ex]].
  %%CITATION = doi:10.1103/PhysRevLett.115.222501;%%
  %16 citations counted in INSPIRE as of 25 Oct 2016

%\cite{Ahn:2013poa}
\bibitem{Ahn:2013poa}
  J.~K.~Ahn {\it et al.} [E373 (KEK-PS) Collaboration],
  %``Double-$\Lambda$ hypernuclei observed in a hybrid emulsion experiment,''
  Phys.\ Rev.\ C {\bf 88}, 014003 (2013).
  %doi:10.1103/PhysRevC.88.014003
  %%CITATION = doi:10.1103/PhysRevC.88.014003;%%
  %33 citations counted in INSPIRE as of 25 Oct 2016

%\cite{Nakazawa:2015joa}
\bibitem{Nakazawa:2015joa}
  K.~Nakazawa {\it et al.},
  %``The first evidence of a deeply bound state of $\Xi^-$-$^{14}$ N system,''
  PTEP {\bf 2015}, 033D02 (2015).
  %doi:10.1093/ptep/ptv008
  %%CITATION = doi:10.1093/ptep/ptv008;%%
  %11 citations counted in INSPIRE as of 25 Oct 2016








%\cite{jparc}
\bibitem{jparc}
  See list of proposals at http://j-parc.jp/researcher/Hadron/en
  /Proposal\_e.html.

%\cite{Hauenstein:2016zys}
\bibitem{Hauenstein:2016zys}
  F.~Hauenstein {\it et al.} [COSY-TOF Collaboration],
  %``First Model-Independent Measurement of the Spin Triplet $p\Lambda$ Scattering Length from Final State Interaction in the $\vec{p}p \rightarrow pK^{+}\Lambda$ Reaction,''
  arXiv:1607.04783 [nucl-ex].
  %%CITATION = ARXIV:1607.04783;%%
  %2 citations counted in INSPIRE as of 25 Oct 2016



%\cite{Hiyama:2014cua}
\bibitem{Hiyama:2014cua}
  E.~Hiyama, S.~Ohnishi, B.~F.~Gibson and T.~A.~Rijken,
  %``Three-body structure of the $nn\Lambda$ system with $\Lambda N-\Sigma N$ coupling,''
  Phys.\ Rev.\ C {\bf 89}, 061302 (2014)
  %doi:10.1103/PhysRevC.89.061302
  [arXiv:1405.2365 [nucl-th]].
  %%CITATION = doi:10.1103/PhysRevC.89.061302;%%
  %17 citations counted in INSPIRE as of 25 Oct 2016

%\cite{Zhou:2007zze}
\bibitem{Zhou:2007zze}
  X.~R.~Zhou, H.-J.~Schulze, H.~Sagawa, C.~X.~Wu and E.~G.~Zhao,
  %``Hypernuclei in the deformed Skyrme-Hartree-Fock approach,''
  Phys.\ Rev.\ C {\bf 76}, 034312 (2007).
  %doi:10.1103/PhysRevC.76.034312
  %%CITATION = doi:10.1103/PhysRevC.76.034312;%%
  %51 citations counted in INSPIRE as of 25 Oct 2016

%\cite{Massot:2012pf}
\bibitem{Massot:2012pf}
  E.~Massot, J.~Margueron and G.~Chanfray,
  %``On the maximum mass of hyperonic neutron stars,''
  Europhys.\ Lett.\  {\bf 97}, 39002 (2012)
  %doi:10.1209/0295-5075/97/39002
  [arXiv:1201.2772 [nucl-th]].
  %%CITATION = doi:10.1209/0295-5075/97/39002;%%
  %31 citations counted in INSPIRE as of 25 Oct 2016

%\cite{Schulze:2011zza}
\bibitem{Schulze:2011zza}
  H.-J.~Schulze and T.~Rijken,
  %``Maximum mass of hyperon stars with the Nijmegen ES C-08 model,''
  Phys.\ Rev.\ C {\bf 84}, 035801 (2011).
  %doi:10.1103/PhysRevC.84.035801
  %%CITATION = doi:10.1103/PhysRevC.84.035801;%%
  %87 citations counted in INSPIRE as of 25 Oct 2016

%\cite{Hu:2013tma}
\bibitem{Hu:2013tma}
  J.~N.~Hu, A.~Li, H.~Toki and W.~Zuo,
  %``Extended quark mean-field model for neutron stars,''
  Phys.\ Rev.\ C {\bf 89}, 025802 (2014)
  %doi:10.1103/PhysRevC.89.025802
  [arXiv:1307.4154 [nucl-th]].
  %%CITATION = doi:10.1103/PhysRevC.89.025802;%%
  %7 citations counted in INSPIRE as of 25 Oct 2016

%\cite{Miyatsu:2013hea}
\bibitem{Miyatsu:2013hea}
  T.~Miyatsu, S.~Yamamuro and K.~Nakazato,
  %``A new equation of state for neutron star matter with nuclei in the crust and hyperons in the core,''
  Astrophys.\ J.\  {\bf 777}, 4 (2013)
  %doi:10.1088/0004-637X/777/1/4
  [arXiv:1308.6121 [astro-ph.HE]].
  %%CITATION = doi:10.1088/0004-637X/777/1/4;%%
  %25 citations counted in INSPIRE as of 25 Oct 2016

%\cite{Mallick:2012wb}
\bibitem{Mallick:2012wb}
  R.~Mallick,
  %``Maximum mass of a hybrid star having a mixed-phase region based on constraints set by the pulsar PSR J1614-2230,''
  Phys.\ Rev.\ C {\bf 87}, 025804 (2013)
  %doi:10.1103/PhysRevC.87.025804
  [arXiv:1207.4872 [astro-ph.HE]].
  %%CITATION = doi:10.1103/PhysRevC.87.025804;%%
  %13 citations counted in INSPIRE as of 25 Oct 2016


%\cite{Demorest:2010bx}
\bibitem{Demorest:2010bx}
  P.~Demorest, T.~Pennucci, S.~Ransom, M.~Roberts and J.~Hessels,
  %``Shapiro Delay Measurement of A Two Solar Mass Neutron Star,''
  Nature {\bf 467}, 1081 (2010)
  %doi:10.1038/nature09466
  [arXiv:1010.5788 [astro-ph.HE]].
  %%CITATION = doi:10.1038/nature09466;%%
  %1270 citations counted in INSPIRE as of 25 Oct 2016

%\cite{Antoniadis:2013pzd}
\bibitem{Antoniadis:2013pzd}
  J.~Antoniadis {\it et al.},
  %``A Massive Pulsar in a Compact Relativistic Binary,''
  Science {\bf 340}, 6131 (2013)
  %doi:10.1126/science.1233232
  [arXiv:1304.6875 [astro-ph.HE]].
  %%CITATION = doi:10.1126/science.1233232;%%
  %422 citations counted in INSPIRE as of 25 Oct 2016

%\cite{Nagels:1976xq}
\bibitem{Nagels:1976xq}
  M.~M.~Nagels, T.~A.~Rijken and J.~J.~de Swart,
  %``Baryon Baryon Scattering in a One Boson Exchange Potential Approach. 2. Hyperon-Nucleon Scattering,''
  Phys.\ Rev.\ D {\bf 15}, 2547 (1977).
  %doi:10.1103/PhysRevD.15.2547
  %%CITATION = doi:10.1103/PhysRevD.15.2547;%%
  %391 citations counted in INSPIRE as of 06 Jan 2016


%\cite{Maessen:1989sx}
\bibitem{Maessen:1989sx}
  P.~M.~M.~Maessen, T.~A.~Rijken and J.~J.~de Swart,
  %``Soft Core Baryon Baryon One Boson Exchange Models. 2. Hyperon - Nucleon Potential,''
  Phys.\ Rev.\ C {\bf 40}, 2226 (1989).
  %doi:10.1103/PhysRevC.40.2226
  %%CITATION = doi:10.1103/PhysRevC.40.2226;%%
  %332 citations counted in INSPIRE as of 06 Jan 2016

%\cite{Rijken:1998yy}
\bibitem{Rijken:1998yy}
  T.~A.~Rijken, V.~G.~J.~Stoks and Y.~Yamamoto,
  %``Soft core hyperon - nucleon potentials,''
  Phys.\ Rev.\ C {\bf 59}, 21 (1999)
  %doi:10.1103/PhysRevC.59.21
  [nucl-th/9807082].
  %%CITATION = doi:10.1103/PhysRevC.59.21;%%
  %363 citations counted in INSPIRE as of 25 Dec 2015

%\cite{Rijken:2006ep}
\bibitem{Rijken:2006ep}
  T.~A.~Rijken and Y.~Yamamoto,
  %``Extended-soft-core baryon-baryon model. II. Hyperon-nucleon interaction,''
  Phys.\ Rev.\ C {\bf 73}, 044008 (2006)
  %doi:10.1103/PhysRevC.73.044008
  [nucl-th/0603042].
  %%CITATION = doi:10.1103/PhysRevC.73.044008;%%
  %112 citations counted in INSPIRE as of 06 Jan 2016

%\cite{Nagels:2015lfa}
\bibitem{Nagels:2015lfa}
  M.~M.~Nagels, T.~A.~Rijken and Y.~Yamamoto,
  %``Extended-soft-core Baryon-Baryon Model Esc08 II. Hyperon-Nucleon Interactions,''
  arXiv:1501.06636 [nucl-th].
  %%CITATION = ARXIV:1501.06636;%%
  %3 citations counted in INSPIRE as of 06 Jan 2016

%\cite{Holzenkamp:1989tq}
\bibitem{Holzenkamp:1989tq}
  B.~Holzenkamp, K.~Holinde and J.~Speth,
  %``A Meson Exchange Model for the Hyperon Nucleon Interaction,''
  Nucl.\ Phys.\ A {\bf 500}, 485 (1989).
  %doi:10.1016/0375-9474(89)90223-6
  %%CITATION = doi:10.1016/0375-9474(89)90223-6;%%
  %287 citations counted in INSPIRE as of 25 Dec 2015

%\cite{Reuber:1993ip}
\bibitem{Reuber:1993ip}
  A.~Reuber, K.~Holinde and J.~Speth,
  %``Meson exchange hyperon - nucleon interactions in free scattering and nuclear matter,''
  Nucl.\ Phys.\ A {\bf 570}, 543 (1994).
  %doi:10.1016/0375-9474(94)90073-6
  %%CITATION = doi:10.1016/0375-9474(94)90073-6;%%
  %159 citations counted in INSPIRE as of 12 Nov 2016

%\cite{Haidenbauer:2005zh}
\bibitem{Haidenbauer:2005zh}
  J.~Haidenbauer and U.~-G.~Mei\ss ner,
  %``The Julich hyperon-nucleon model revisited,''
  Phys.\ Rev.\ C {\bf 72}, 044005 (2005)
  %doi:10.1103/PhysRevC.72.044005
  [nucl-th/0506019].
  %%CITATION = doi:10.1103/PhysRevC.72.044005;%%
  %90 citations counted in INSPIRE as of 06 Jan 2016

%\cite{Straub:1988gj}
\bibitem{Straub:1988gj}
  U.~Straub, Z.~Y.~Zhang, K.~Brauer, A.~Faessler, S.~B.~Khadkikar and G.~Lubeck,
  %``Hyperon Nucleon Interaction in the Quark Cluster Model,''
  Nucl.\ Phys.\ A {\bf 483}, 686 (1988).
 % doi:10.1016/0375-9474(88)90092-9
  %%CITATION = doi:10.1016/0375-9474(88)90092-9;%%
  %62 citations counted in INSPIRE as of 06 Jan 2016

%\cite{Straub:1990de}
\bibitem{Straub:1990de}
  U.~Straub, Z.~Y.~Zhang, K.~Braeuer, A.~Faessler, S.~B.~Khadkikar and G.~Luebeck,
  %``Hyperon nucleon interaction and the H dibaryon in the quark cluster model,''
  Nucl.\ Phys.\ A {\bf 508}, 385C (1990).
  %doi:10.1016/0375-9474(90)90502-D
  %%CITATION = doi:10.1016/0375-9474(90)90502-D;%%
  %33 citations counted in INSPIRE as of 06 Jan 2016

%\cite{Zhang:1994pp}
\bibitem{Zhang:1994pp}
  Z.~Y.~Zhang, A.~Faessler, U.~Straub and L.~Y.~Glozman,
  %``The Baryon baryon interaction in a modified quark model,''
  Nucl.\ Phys.\ A {\bf 578}, 573 (1994).
  %doi:10.1016/0375-9474(94)90761-7
  %%CITATION = doi:10.1016/0375-9474(94)90761-7;%%
  %56 citations counted in INSPIRE as of 01 fÅ¡Å vr. 2016

%\cite{Zhang:1997ny}
\bibitem{Zhang:1997ny}
  Z.~Y.~Zhang, Y.~W.~Yu, P.~N.~Shen, L.~R.~Dai, A.~Faessler and U.~Straub,
  %``Hyperon nucleon interactions in a chiral SU(3) quark model,''
  Nucl.\ Phys.\ A {\bf 625}, 59 (1997).
  %doi:10.1016/S0375-9474(97)00033-X
  %%CITATION = doi:10.1016/S0375-9474(97)00033-X;%%
  %106 citations counted in INSPIRE as of 16 fÅ¡Å vr. 2016

%\cite{Ping:1998si}
\bibitem{Ping:1998si}
  J.~L.~Ping, F.~Wang and J.~T.~Goldman,
  %``Effective baryon baryon potentials in the quark delocalization and color screening model,''
  Nucl.\ Phys.\ A {\bf 657}, 95 (1999)
  %doi:10.1016/S0375-9474(99)00321-8
  [nucl-th/9812068].
  %%CITATION = doi:10.1016/S0375-9474(99)00321-8;%%
  %35 citations counted in INSPIRE as of 01 fÅ¡Å vr. 2016

%\cite{Fujiwara:1995fx}
\bibitem{Fujiwara:1995fx}
  Y.~Fujiwara, C.~Nakamoto and Y.~Suzuki,
  %``A Unified description of N N and Y N interactions in a quark model with effective meson exchange potentials,''
  Phys.\ Rev.\ Lett.\  {\bf 76}, 2242 (1996).
  %doi:10.1103/PhysRevLett.76.2242
  %%CITATION = doi:10.1103/PhysRevLett.76.2242;%%
  %97 citations counted in INSPIRE as of 06 Jan 2016

%\cite{Fujiwara:2006yh}
\bibitem{Fujiwara:2006yh}
  Y.~Fujiwara, Y.~Suzuki and C.~Nakamoto,
  %``Baryon-baryon interactions in the SU(6) quark model and their applications to light nuclear systems,''
  Prog.\ Part.\ Nucl.\ Phys.\  {\bf 58}, 439 (2007)
  %doi:10.1016/j.ppnp.2006.08.001
  [nucl-th/0607013].
  %%CITATION = doi:10.1016/j.ppnp.2006.08.001;%%
  %108 citations counted in INSPIRE as of 16 fÅ¡Å vr. 2016


%\cite{Beane:2006gf}
\bibitem{Beane:2006gf}
  S.~R.~Beane {\it et al.} [NPLQCD Collaboration],
  %``Hyperon-Nucleon Scattering from Fully-Dynamical Lattice QCD,''
  Nucl.\ Phys.\ A {\bf 794}, 62 (2007)
  %doi:10.1016/j.nuclphysa.2007.07.006
  [hep-lat/0612026].
  %%CITATION = doi:10.1016/j.nuclphysa.2007.07.006;%%
  %80 citations counted in INSPIRE as of 06 Jan 2016

%\cite{Nemura:2008sp}
\bibitem{Nemura:2008sp}
  H.~Nemura, N.~Ishii, S.~Aoki and T.~Hatsuda,
  %``Hyperon-nucleon force from lattice QCD,''
  Phys.\ Lett.\ B {\bf 673}, 136 (2009)
  %doi:10.1016/j.physletb.2009.02.003
  [arXiv:0806.1094 [nucl-th]].
  %%CITATION = doi:10.1016/j.physletb.2009.02.003;%%
  %97 citations counted in INSPIRE as of 06 Jan 2016

%\cite{Beane:2009py}
\bibitem{Beane:2009py}
  S.~R.~Beane {\it et al.} [NPLQCD Collaboration],
  %``High Statistics Analysis using Anisotropic Clover Lattices: (III) Baryon-Baryon Interactions,''
  Phys.\ Rev.\ D {\bf 81}, 054505 (2010)
  %doi:10.1103/PhysRevD.81.054505
  [arXiv:0912.4243 [hep-lat]].
  %%CITATION = doi:10.1103/PhysRevD.81.054505;%%
  %63 citations counted in INSPIRE as of 01 fÅ¡Å vr. 2016

%\cite{Inoue:2010hs}
\bibitem{Inoue:2010hs}
  T.~Inoue {\it et al.} [HAL QCD Collaboration],
  %``Baryon-Baryon Interactions in the Flavor SU(3) Limit from Full QCD Simulations on the Lattice,''
  Prog.\ Theor.\ Phys.\  {\bf 124}, 591 (2010)
  %doi:10.1143/PTP.124.591
  [arXiv:1007.3559 [hep-lat]].
  %%CITATION = doi:10.1143/PTP.124.591;%%
  %74 citations counted in INSPIRE as of 06 Jan 2016

%\cite{Beane:2011iw}
\bibitem{Beane:2011iw}
  S.~R.~Beane {\it et al.} [NPLQCD Collaboration],
  %``The Deuteron and Exotic Two-Body Bound States from Lattice QCD,''
  Phys.\ Rev.\ D {\bf 85}, 054511 (2012)
  %doi:10.1103/PhysRevD.85.054511
  [arXiv:1109.2889 [hep-lat]].
  %%CITATION = doi:10.1103/PhysRevD.85.054511;%%
  %94 citations counted in INSPIRE as of 01 Feb 2016

%\cite{Sasaki:2015ifa}
\bibitem{Sasaki:2015ifa}
  K.~Sasaki {\it et al.} [HAL QCD Collaboration],
  %``Coupled-channel approach to strangeness S?=??2 baryonÅ¡Cbayron interactions in lattice QCD,''
  PTEP {\bf 2015}, 113B01 (2015)
  %doi:10.1093/ptep/ptv144
  [arXiv:1504.01717 [hep-lat]].
  %%CITATION = doi:10.1093/ptep/ptv144;%%
  %3 citations counted in INSPIRE as of 01 fÅ¡Å vr. 2016

%\cite{Doi:2015oha}
\bibitem{Doi:2015oha}
  T.~Doi {\it et al.},
  %``First results of baryon interactions from lattice QCD with physical masses (1) -- General overview and two-nucleon forces --,''
  arXiv:1512.01610 [hep-lat].
  %%CITATION = ARXIV:1512.01610;%%
  %1 citations counted in INSPIRE as of 17 fÅ¡Å vr. 2016

%\cite{Doi:2015uvd}
\bibitem{Doi:2015uvd}
  T.~Doi {\it et al.},
  %``Towards Lattice QCD Baryon Forces at the Physical Point: First Results,''
  arXiv:1512.04199 [hep-lat].
  %%CITATION = ARXIV:1512.04199;%%


%\cite{Bedaque:2002mn}
\bibitem{Bedaque:2002mn}
  P.~F.~Bedaque and U.~van Kolck,
  %``Effective field theory for few nucleon systems,''
  Ann.\ Rev.\ Nucl.\ Part.\ Sci.\  {\bf 52}, 339 (2002)
  %doi:10.1146/annurev.nucl.52.050102.090637
  [nucl-th/0203055].
  %%CITATION = doi:10.1146/annurev.nucl.52.050102.090637;%%
  %455 citations counted in INSPIRE as of 19 Mar 2016

%\cite{Epelbaum:2008ga}
\bibitem{Epelbaum:2008ga}
  E.~Epelbaum, H.~W.~Hammer and U.~-G.~Mei\ss ner,
  %``Modern Theory of Nuclear Forces,''
  Rev.\ Mod.\ Phys.\  {\bf 81}, 1773 (2009)
  %doi:10.1103/RevModPhys.81.1773
  [arXiv:0811.1338 [nucl-th]].
  %%CITATION = doi:10.1103/RevModPhys.81.1773;%%
  %548 citations counted in INSPIRE as of 06 Jan 2016

%\cite{Machleidt:2011zz}
\bibitem{Machleidt:2011zz}
  R.~Machleidt and D.~R.~Entem,
  %``Chiral effective field theory and nuclear forces,''
  Phys.\ Rept.\  {\bf 503}, 1 (2011)
  %doi:10.1016/j.physrep.2011.02.001
  [arXiv:1105.2919 [nucl-th]].
  %%CITATION = doi:10.1016/j.physrep.2011.02.001;%%
  %340 citations counted in INSPIRE as of 06 Jan 2016

%\cite{Weinberg:1990rz}
\bibitem{Weinberg:1990rz}
  S.~Weinberg,
  %``Nuclear forces from chiral Lagrangians,''
  Phys.\ Lett.\ B {\bf 251}, 288 (1990).
  %doi:10.1016/0370-2693(90)90938-3
  %%CITATION = doi:10.1016/0370-2693(90)90938-3;%%
  %963 citations counted in INSPIRE as of 04 Mar 2016


%\cite{Weinberg:1991um}
\bibitem{Weinberg:1991um}
  S.~Weinberg,
  %``Effective chiral Lagrangians for nucleon - pion interactions and nuclear forces,''
  Nucl.\ Phys.\ B {\bf 363}, 3 (1991).
  %doi:10.1016/0550-3213(91)90231-L
  %%CITATION = doi:10.1016/0550-3213(91)90231-L;%%
  %929 citations counted in INSPIRE as of 04 Mar 2016

%\cite{Kang:2013uia}
\bibitem{Kang:2013uia}
  X.~W.~Kang, J.~Haidenbauer and U.~-G.~Mei\ss ner,
  %``Antinucleon-nucleon interaction in chiral effective field theory,''
  JHEP {\bf 1402}, 113 (2014)
  %doi:10.1007/JHEP02(2014)113
  [arXiv:1311.1658 [hep-ph]].
  %%CITATION = doi:10.1007/JHEP02(2014)113;%%
  %11 citations counted in INSPIRE as of 01 avril 2016

%\cite{Polinder:2006zh}
\bibitem{Polinder:2006zh}
  H.~Polinder, J.~Haidenbauer and U.~-G.~Mei\ss ner,
  %``Hyperon-nucleon interactions: A Chiral effective field theory approach,''
  Nucl.\ Phys.\ A {\bf 779}, 244 (2006)
  %doi:10.1016/j.nuclphysa.2006.09.006
  [nucl-th/0605050].
  %%CITATION = doi:10.1016/j.nuclphysa.2006.09.006;%%
  %82 citations counted in INSPIRE as of 25 Dec 2015

%\cite{Haidenbauer:2007ra}
\bibitem{Haidenbauer:2007ra}
  J.~Haidenbauer, U.~-G.~Mei\ss ner, A.~Nogga and H.~Polinder,
  %``The Hyperon-nucleon interaction: Conventional versus effective field theory approach,''
  Lect.\ Notes Phys.\  {\bf 724}, 113 (2007)
  %doi:10.1007/978-3-540-72039-3\_4
  [nucl-th/0702015 [NUCL-TH]].
  %%CITATION = doi:10.1007/978-3-540-72039-3_4;%%
  %37 citations counted in INSPIRE as of 06 Jan 2016

%\cite{Haidenbauer:2013oca}
\bibitem{Haidenbauer:2013oca}
  J.~Haidenbauer, S.~Petschauer, N.~Kaiser, U.-G.~Mei\ss ner, A.~Nogga and W.~Weise,
  %``Hyperon-nucleon interaction at next-to-leading order in chiral effective field theory,''
  Nucl.\ Phys.\ A {\bf 915}, 24 (2013)
  %doi:10.1016/j.nuclphysa.2013.06.008
  [arXiv:1304.5339 [nucl-th]].
  %%CITATION = doi:10.1016/j.nuclphysa.2013.06.008;%%
  %41 citations counted in INSPIRE as of 25 Dec 2015

%\cite{Polinder:2007mp}
\bibitem{Polinder:2007mp}
  H.~Polinder, J.~Haidenbauer and U.-G.~Mei\ss ner,
  %``Strangeness S = -2 baryon-baryon interactions using chiral effective field theory,''
  Phys.\ Lett.\ B {\bf 653}, 29 (2007)
  %doi:10.1016/j.physletb.2007.07.045
  [arXiv:0705.3753 [nucl-th]].
  %%CITATION = doi:10.1016/j.physletb.2007.07.045;%%
  %56 citations counted in INSPIRE as of 15  . 2016


%\cite{Haidenbauer:2009qn}
\bibitem{Haidenbauer:2009qn}
  J.~Haidenbauer and U.-G.~Mei\ss ner,
  %``Predictions for the strangeness S = -3 and -4 baryon-baryon interactions in chiral effective field theory,''
  Phys.\ Lett.\ B {\bf 684}, 275 (2010)
  %doi:10.1016/j.physletb.2010.01.031
  [arXiv:0907.1395 [nucl-th]].
  %%CITATION = doi:10.1016/j.physletb.2010.01.031;%%
  %29 citations counted in INSPIRE as of 15  . 2016

%\cite{Haidenbauer:2015zqb}
\bibitem{Haidenbauer:2015zqb}
  J.~Haidenbauer, U.-G.~Mei\ss ner and S.~Petschauer,
  %``Strangeness S = -2 baryonšCbaryon interaction at next-to-leading order in chiral effective field theory,''
  Nucl.\ Phys.\ A {\bf 954}, 273 (2016)
  %doi:10.1016/j.nuclphysa.2016.01.006
  [arXiv:1511.05859 [nucl-th]].
  %%CITATION = doi:10.1016/j.nuclphysa.2016.01.006;%%
  %11 citations counted in INSPIRE as of 26 Oct 2016






%\cite{Lepage:1997cs}
\bibitem{Lepage:1997cs}
  G.~P.~Lepage,
  %``How to renormalize the Schrodinger equation,''
  nucl-th/9706029.
  %%CITATION = NUCL-TH/9706029;%%
  %319 citations counted in INSPIRE as of 01 fÅ¡Å vr. 2016

%\cite{Birse:2005um}
\bibitem{Birse:2005um}
  M.~C.~Birse,
  %``Power counting with one-pion exchange,''
  Phys.\ Rev.\ C {\bf 74}, 014003 (2006)
  %doi:10.1103/PhysRevC.74.014003
  [nucl-th/0507077].
  %%CITATION = doi:10.1103/PhysRevC.74.014003;%%
  %111 citations counted in INSPIRE as of 19 Mar 2016

%\cite{Nogga:2005hy}
\bibitem{Nogga:2005hy}
  A.~Nogga, R.~G.~E.~Timmermans and U.~van Kolck,
  %``Renormalization of one-pion exchange and power counting,''
  Phys.\ Rev.\ C {\bf 72}, 054006 (2005)
  %doi:10.1103/PhysRevC.72.054006
  [nucl-th/0506005].
  %%CITATION = doi:10.1103/PhysRevC.72.054006;%%
  %197 citations counted in INSPIRE as of 01 fÅ¡Å vr. 2016

%\cite{Epelbaum:2006pt}
\bibitem{Epelbaum:2006pt}
  E.~Epelbaum and U.-G.~Mei\ss ner,
  %``On the Renormalization of the One-Pion Exchange Potential and the Consistency of Weinberg's's`s Power Counting,''
  Few Body Syst.\  {\bf 54}, 2175 (2013)
  %doi:10.1007/s00601-012-0492-1
  [nucl-th/0609037].
  %%CITATION = doi:10.1007/s00601-012-0492-1;%%
  %101 citations counted in INSPIRE as of 01 Feb 2016

%\cite{Long:2007vp}
\bibitem{Long:2007vp}
  B.~Long and U.~van Kolck,
  %``Renormalization of Singular Potentials and Power Counting,''
  Annals Phys.\  {\bf 323}, 1304 (2008)
  %doi:10.1016/j.aop.2008.01.003
  [arXiv:0707.4325 [quant-ph]].
  %%CITATION = doi:10.1016/j.aop.2008.01.003;%%
  %38 citations counted in INSPIRE as of 01 fÅ¡Å vr. 2016

%\cite{Yang:2009pn}
\bibitem{Yang:2009pn}
  C.-J.~Yang, C.~Elster and D.~R.~Phillips,
  %``Subtractive renormalization of the NN interaction in chiral effective theory up to next-to-next-to-leading order: S waves,''
  Phys.\ Rev.\ C {\bf 80}, 044002 (2009)
  %doi:10.1103/PhysRevC.80.044002
  [arXiv:0905.4943 [nucl-th]].
  %%CITATION = doi:10.1103/PhysRevC.80.044002;%%
  %34 citations counted in INSPIRE as of 01 fÅ¡Å vr. 2016

%\cite{Valderrama:2009ei}
\bibitem{Valderrama:2009ei}
  M.~P.~Valderrama,
  %``Perturbative renormalizability of chiral two pion exchange in nucleon-nucleon scattering,''
  Phys.\ Rev.\ C {\bf 83}, 024003 (2011)
  %doi:10.1103/PhysRevC.83.024003
  [arXiv:0912.0699 [nucl-th]].
  %%CITATION = doi:10.1103/PhysRevC.83.024003;%%
  %52 citations counted in INSPIRE as of 01 fÅ¡Å vr. 2016

%\cite{Long:2011xw}
\bibitem{Long:2011xw}
  B.~Long and C.~J.~Yang,
  %``Renormalizing Chiral Nuclear Forces: Triplet Channels,''
  Phys.\ Rev.\ C {\bf 85}, 034002 (2012)
  %doi:10.1103/PhysRevC.85.034002
  [arXiv:1111.3993 [nucl-th]].
  %%CITATION = doi:10.1103/PhysRevC.85.034002;%%
  %30 citations counted in INSPIRE as of 01 fÅ¡Å vr. 2016

















%\cite{Epelbaum:2012ua}
\bibitem{Epelbaum:2012ua}
  E.~Epelbaum and J.~Gegelia,
  %``Weinberg's's's approach to nucleon?Cnucleon scattering revisited,''
  Phys.\ Lett.\ B {\bf 716}, 338 (2012)
  %doi:10.1016/j.physletb.2012.08.025
  [arXiv:1207.2420 [nucl-th]].
  %%CITATION = doi:10.1016/j.physletb.2012.08.025;%%
  %27 citations counted in INSPIRE as of 03 Jan 2016


%\cite{Epelbaum:2015sha}
\bibitem{Epelbaum:2015sha}
  E.~Epelbaum, A.~M.~Gasparyan, J.~Gegelia and H.~Krebs,
  %``$^{1}$S$_{0}$ nucleon-nucleon scattering in the modified Weinberg's's approach,''
  Eur.\ Phys.\ J.\ A {\bf 51}, 71 (2015)
  %doi:10.1140/epja/i2015-15071-6
  [arXiv:1501.01191 [nucl-th]].
  %%CITATION = doi:10.1140/epja/i2015-15071-6;%%
  %2 citations counted in INSPIRE as of 01 Feb 2016

%\cite{Stoks:1993tb}
\bibitem{Stoks:1993tb}
  V.~G.~J.~Stoks, R.~A.~M.~Klomp, M.~C.~M.~Rentmeester and J.~J.~de Swart,
  %``Partial wave analaysis of all nucleon-nucleon scattering data below 350-MeV,''
  Phys.\ Rev.\ C {\bf 48}, 792 (1993).
  %doi:10.1103/PhysRevC.48.792
  %%CITATION = doi:10.1103/PhysRevC.48.792;%%
  %646 citations counted in INSPIRE as of 12 Dec 2016

%\cite{Li:2016paq}
\bibitem{Li:2016paq}
  K.~-W.~Li, X.~-L.~Ren, L.~S.~Geng and B.~Long,
  %``Strangeness $S=-1$ hyperon-nucleon scattering in covariant chiral effective field theory,''
  Phys.\ Rev.\ D {\bf 94}, 014029 (2016)
  %doi:10.1103/PhysRevD.94.014029
  [arXiv:1603.07802 [hep-ph]].
  %%CITATION = doi:10.1103/PhysRevD.94.014029;%%
  %1 citations counted in INSPIRE as of 03 Nov 2016


  %\cite{Geng:2008mf}
\bibitem{Geng:2008mf}
  L.~S.~Geng, J.~Martin Camalich, L.~Alvarez-Ruso and M.~J.~Vicente Vacas,
  %``Leading SU(3)-breaking corrections to the baryon magnetic moments in Chiral Perturbation Theory,''
  Phys.\ Rev.\ Lett.\  {\bf 101}, 222002 (2008)
  %doi:10.1103/PhysRevLett.101.222002
  [arXiv:0805.1419 [hep-ph]].
  %%CITATION = doi:10.1103/PhysRevLett.101.222002;%%
  %64 citations counted in INSPIRE as of 01 fÅ¡Å vr. 2016

  %\cite{Geng:2009ik}
\bibitem{Geng:2009ik}
  L.~S.~Geng, J.~Martin Camalich and M.~J.~Vicente Vacas,
  %``SU(3)-breaking corrections to the hyperon vector coupling f(1)(0) in covariant baryon chiral perturbation theory,''
  Phys.\ Rev.\ D {\bf 79}, 094022 (2009)
  %doi:10.1103/PhysRevD.79.094022
  [arXiv:0903.4869 [hep-ph]].
  %%CITATION = doi:10.1103/PhysRevD.79.094022;%%
  %28 citations counted in INSPIRE as of 01 fÅ


  %\cite{Geng:2011wq}
\bibitem{Geng:2011wq}
  L.~S.~Geng, X.~-L.~Ren, J.~Martin-Camalich and W.~Weise,
  %``Finite-volume effects on octet-baryon masses in covariant baryon chiral perturbation theory,''
  Phys.\ Rev.\ D {\bf 84}, 074024 (2011)
  %doi:10.1103/PhysRevD.84.074024
  [arXiv:1108.2231 [hep-ph]].
  %%CITATION = doi:10.1103/PhysRevD.84.074024;%%
  %21 citations counted in INSPIRE as of 01 fÅ

  %\cite{Ren:2012aj}
\bibitem{Ren:2012aj}
  X.~-L.~Ren, L.~S.~Geng, J.~Martin Camalich, J.~Meng and H.~Toki,
  %``Octet baryon masses in next-to-next-to-next-to-leading order covariant baryon chiral perturbation theory,''
  JHEP {\bf 1212}, 073 (2012)
  %doi:10.1007/JHEP12(2012)073
  [arXiv:1209.3641 [nucl-th]].
  %%CITATION = doi:10.1007/JHEP12(2012)073;%%
  %36 citations counted in INSPIRE as of 01 fÅ

  %\cite{Ren:2014vea}
\bibitem{Ren:2014vea}
  X.~-L.~Ren, L.~-S.~Geng and J.~Meng,
  %``Scalar strangeness content of the nucleon and baryon sigma terms,''
  Phys.\ Rev.\ D {\bf 91}, 051502 (2015)
  %doi:10.1103/PhysRevD.91.051502
  [arXiv:1404.4799 [hep-ph]].
  %%CITATION = doi:10.1103/PhysRevD.91.051502;%%
  %17 citations counted in INSPIRE as of 01 fÅ

 %\cite{Geng:2010vw}
\bibitem{Geng:2010vw}
  L.~S.~Geng, N.~Kaiser, J.~Martin-Camalich and W.~Weise,
  %``Low-energy interactions of Nambu-Goldstone bosons with $D$ mesons in covariant chiral perturbation theory,''
  Phys.\ Rev.\ D {\bf 82}, 054022 (2010)
  %doi:10.1103/PhysRevD.82.054022
  [arXiv:1008.0383 [hep-ph]].
  %%CITATION = doi:10.1103/PhysRevD.82.054022;%%
  %30 citations counted in INSPIRE as of 01 Feb 2016


  %\cite{Geng:2010df}
\bibitem{Geng:2010df}
  L.~S.~Geng, M.~Altenbuchinger and W.~Weise,
  %``Light quark mass dependence of the $D$ and $D_s$ decay constants,''
  Phys.\ Lett.\ B {\bf 696}, 390 (2011)
  %doi:10.1016/j.physletb.2010.12.060
  [arXiv:1012.0666 [hep-ph]].
  %%CITATION = doi:10.1016/j.physletb.2010.12.060;%%
  %11 citations counted in INSPIRE as of 01 fÅ

 %\cite{Altenbuchinger:2011qn}
\bibitem{Altenbuchinger:2011qn}
  M.~Altenbuchinger, L.~S.~Geng and W.~Weise,
  %``SU(3) Breaking Corrections to the $D$, $D^*$, $B$, and $B^*$ decay Constants,''
  Phys.\ Lett.\ B {\bf 713}, 453 (2012)
  %doi:10.1016/j.physletb.2012.06.025
  [arXiv:1109.0460 [hep-ph]].
  %%CITATION = doi:10.1016/j.physletb.2012.06.025;%%



%\cite{Geng:2013xn}
\bibitem{Geng:2013xn}
  L.~S.~Geng,
  %``Recent developments in SU(3) covariant baryon chiral perturbation theory,''
  Front.\ Phys.\ (Beijing) {\bf 8}, 328 (2013)
  %doi:10.1007/s11467-013-0327-7
  [arXiv:1301.6815 [nucl-th]].
  %%CITATION = doi:10.1007/s11467-013-0327-7;%%
  %19 citations counted in INSPIRE as of 12 Dec 2016


%\cite{Ren:2016jna}
\bibitem{Ren:2016jna} 
  X.~L.~Ren, K.~W.~Li, L.~S.~Geng, B.~W.~Long, P.~Ring and J.~Meng,
  %``Leading order relativistic chiral nucleon-nucleon interaction,''
  Chin. Phys. C  42, 014103(2018)
  %doi:10.1088/1674-1137/42/1/014103
  [arXiv:1611.08475 [nucl-th]].
  %%CITATION = doi:10.1088/1674-1137/42/1/014103;%%
  %10 citations counted in INSPIRE as of 04 Jan 2018

%\cite{Girlanda:2010ya}
\bibitem{Girlanda:2010ya}
  L.~Girlanda, S.~Pastore, R.~Schiavilla and M.~Viviani,
  %``Relativity constraints on the two-nucleon contact interaction,''
  Phys.\ Rev.\ C {\bf 81}, 034005 (2010)
  %doi:10.1103/PhysRevC.81.034005
  [arXiv:1001.3676 [nucl-th]].
  %%CITATION = doi:10.1103/PhysRevC.81.034005;%%
  %25 citations counted in INSPIRE as of 25 Sep 2017

%\cite{Djukanovic:2007zz}
\bibitem{Djukanovic:2007zz}
  D.~Djukanovic, J.~Gegelia, S.~Scherer and M.~R.~Schindler,
  %``N N scattering in higher-derivative formulation of baryon chiral perturbation theory,''
  Few Body Syst.\  {\bf 41}, 141 (2007)
  %doi:10.1007/s00601-007-0194-2
  [nucl-th/0609055].
  %%CITATION = doi:10.1007/s00601-007-0194-2;%%
  %13 citations counted in INSPIRE as of 23 May 2017

%%\cite{Girlanda:2010ya}
%\bibitem{Girlanda:2010ya}
%  L.~Girlanda, S.~Pastore, R.~Schiavilla and M.~Viviani,
%  %``Relativity constraints on the two-nucleon contact interaction,''
%  Phys.\ Rev.\ C {\bf 81}, 034005 (2010)
%  %doi:10.1103/PhysRevC.81.034005
%  [arXiv:1001.3676 [nucl-th]].
%  %%CITATION = doi:10.1103/PhysRevC.81.034005;%%
%  %25 citations counted in INSPIRE as of 23 May 2017

%\cite{Petschauer:2013uua}
\bibitem{Petschauer:2013uua}
  S.~Petschauer and N.~Kaiser,
  %``Relativistic SU(3) chiral baryon-baryon Lagrangian up to order $q^{2}$,''
  Nucl.\ Phys.\ A {\bf 916}, 1 (2013)
  %doi:10.1016/j.nuclphysa.2013.07.010
  [arXiv:1305.3427 [nucl-th]].
  %%CITATION = doi:10.1016/j.nuclphysa.2013.07.010;%%
  %14 citations counted in INSPIRE as of 23 May 2017

%\cite{Patrignani:2016xqp}
\bibitem{Patrignani:2016xqp}
  C.~Patrignani {\it et al.} [Particle Data Group],
  %``Review of Particle Physics,''
  Chin.\ Phys.\ C {\bf 40}, 100001 (2016).
  %doi:10.1088/1674-1137/40/10/100001
  %%CITATION = doi:10.1088/1674-1137/40/10/100001;%%
  %1561 citations counted in INSPIRE as of 23 Sep 2017

%\cite{Woloshyn:1974wm}
\bibitem{Woloshyn:1974wm}
  R.~M.~Woloshyn and A.~D.~Jackson,
  %``Comparison of three-dimensional relativistic scattering equations,''
  Nucl.\ Phys.\ B {\bf 64}, 269 (1973).
  %doi:10.1016/0550-3213(73)90626-3
  %%CITATION = doi:10.1016/0550-3213(73)90626-3;%%
  %80 citations counted in INSPIRE as of 03 Jan 2016

%%\cite{Thompson:1970wt}
%\bibitem{Thompson:1970wt}
%  R.~H.~Thompson,
%  %``Three-dimensional bethe-salpeter equation applied to the nucleon-nucleon interaction,''
%  Phys.\ Rev.\ D {\bf 1}, 110 (1970).
%%  doi:10.1103/PhysRevD.1.110
%  %%CITATION = doi:10.1103/PhysRevD.1.110;%%
%  %129 citations counted in INSPIRE as of 12 Jul 2017


%\cite{Epelbaum:2004fk}
\bibitem{Epelbaum:2004fk}
  E.~Epelbaum, W.~Gl\"ockle and U.-G.~Mei\ss ner,
  %``The Two-nucleon system at next-to-next-to-next-to-leading order,''
  Nucl.\ Phys.\ A {\bf 747}, 362 (2005)
  %doi:10.1016/j.nuclphysa.2004.09.107
  [nucl-th/0405048].
  %%CITATION = doi:10.1016/j.nuclphysa.2004.09.107;%%
  %470 citations counted in INSPIRE as of 12 Dec 2016

%\cite{Vincent:1974zz}
\bibitem{Vincent:1974zz}
  C.~M.~Vincent and S.~C.~Phatak,
  %``Accurate momentum-space method for scattering by nuclear and Coulomb potentials,''
  Phys.\ Rev.\ C {\bf 10}, 391 (1974).
  %doi:10.1103/PhysRevC.10.391
  %%CITATION = doi:10.1103/PhysRevC.10.391;%%
  %97 citations counted in INSPIRE as of 25 Dec 2015

%\cite{SechiZorn:1969hk}
\bibitem{SechiZorn:1969hk}
  B.~Sechi-Zorn, B.~Kehoe, J.~Twitty and R.~A.~Burnstein,
  %``Low-energy lambda-proton elastic scattering,''
  Phys.\ Rev.\  {\bf 175}, 1735 (1968).
 % doi:10.1103/PhysRev.175.1735
  %%CITATION = doi:10.1103/PhysRev.175.1735;%%
  %156 citations counted in INSPIRE as of 23 Jun 2016

%\cite{Alexander:1969cx}
\bibitem{Alexander:1969cx}
  G.~Alexander, U.~Karshon, A.~Shapira, G.~Yekutieli, R.~Engelmann, H.~Filthuth and W.~Lughofer,
  %``Study of the lambda-n system in low-energy lambda-p elastic scattering,''
  Phys.\ Rev.\  {\bf 173}, 1452 (1968).
  %doi:10.1103/PhysRev.173.1452
  %%CITATION = doi:10.1103/PhysRev.173.1452;%%
  %151 citations counted in INSPIRE as of 25 Dec 2015

%\cite{Eisele:1971mk}
\bibitem{Eisele:1971mk}
  F.~Eisele, H.~Filthuth, W.~Foehlisch, V.~Hepp and G.~Zech,
  %``Elastic sigma+- p scattering at low energies,''
  Phys.\ Lett.\ B {\bf 37}, 204 (1971).
  %doi:10.1016/0370-2693(71)90053-0
  %%CITATION = doi:10.1016/0370-2693(71)90053-0;%%
  %70 citations counted in INSPIRE as of 25 Dec 2015

%\cite{Engelmann:1966}
\bibitem{Engelmann:1966}
  R.~Engelmann, H.~Filthuth, V.~Hepp,~E. Kluge,
  %``,''
  Phys.\ Lett.\  {\bf 21}, 587 (1966).

  %\cite{Hauptman:1977hr}
\bibitem{Hauptman:1977hr}
  J.~M.~Hauptman, J.~A.~Kadyk and G.~H.~Trilling,
  %``Experimental Study of Lambda p and xi0 p Interactions in the Range 1-GeV/c-10-GeV/c,''
  Nucl.\ Phys.\ B {\bf 125}, 29 (1977).
  %doi:10.1016/0550-3213(77)90222-X
  %%CITATION = doi:10.1016/0550-3213(77)90222-X;%%
  %49 citations counted in INSPIRE as of 23 Mar 2016

  %\cite{Kadyk:1971tc}
\bibitem{Kadyk:1971tc}
  J.~A.~Kadyk, G.~Alexander, J.~H.~Chan, P.~Gaposchkin and G.~H.~Trilling,
  %``Lambda p interactions in momentum range 300 to 1500 mev/c,''
  Nucl.\ Phys.\ B {\bf 27}, 13 (1971).
  %doi:10.1016/0550-3213(71)90076-9
  %%CITATION = doi:10.1016/0550-3213(71)90076-9;%%
  %99 citations counted in INSPIRE as of 23 Mar 2016

%\cite{Hepp:1968zza}
\bibitem{Hepp:1968zza}
  V.~Hepp and H.~Schleich,
  %``A New Determination of the Capture Ratio r(c) = Sigma- p --> Sigma0 n / (Sigma- p --> Sigma0 n) + (Sigma- p --> Lambda n), the Lambda0-Lifetime and Sigma- Lambda0 Mass Difference,''
  Z.\ Phys.\  {\bf 214}, 71 (1968).
  %doi:10.1007/BF01380085
  %%CITATION = doi:10.1007/BF01380085;%%
  %20 citations counted in INSPIRE as of 25 Dec 2015

%\cite{Juric:1973zq}
\bibitem{Juric:1973zq}
  M.~Juric {\it et al.},
  %``A new determination of the binding-energy values of the light hypernuclei (15>=a),''
  Nucl.\ Phys.\ B {\bf 52}, 1 (1973).
  %doi:10.1016/0550-3213(73)90084-9
  %%CITATION = doi:10.1016/0550-3213(73)90084-9;%%
  %192 citations counted in INSPIRE as of 22 Jun 2016


%\cite{Davis:1991zpu}
\bibitem{Davis:1991zpu}
  D.~H.~Davis,
  %``A brief review of emulsion results on hypernuclei,''
  AIP Conf.\ Proc.\  {\bf 224}, 38 (1991).
  %doi:10.1063/1.40526
  %%CITATION = doi:10.1063/1.40526;%%
  %9 citations counted in INSPIRE as of 22 Jun 2016

%\cite{Nogga:2013pwa}
\bibitem{Nogga:2013pwa}
  A.~Nogga,
  %``Light hypernuclei based on chiral and phenomenological interactions,''
  Nucl.\ Phys.\ A {\bf 914}, 140 (2013).
  %doi:10.1016/j.nuclphysa.2013.02.053
  %%CITATION = doi:10.1016/j.nuclphysa.2013.02.053;%%
  %19 citations counted in INSPIRE as of 06 Nov 2016

%%\cite{Korpa:2001au}
%\bibitem{Korpa:2001au}
%  C.~L.~Korpa, A.~E.~L.~Dieperink and R.~G.~E.~Timmermans,
%  %``Hyperon nucleon scattering and hyperon masses in the nuclear medium,''
%  Phys.\ Rev.\ C {\bf 65}, 015208 (2002)
%  %doi:10.1103/PhysRevC.65.015208
%  [nucl-th/0109072].
%  %%CITATION = doi:10.1103/PhysRevC.65.015208;%%
%  %29 citations counted in INSPIRE as of 06 Nov 2016


%\cite{Tominaga:1998iy}
\bibitem{Tominaga:1998iy}
  K.~Tominaga, T.~Ueda, M.~Yamaguchi, N.~Kijima, D.~Okamoto, K.~Miyagawa and T.~Yamada,
  %``A one-boson-exchange potential for Lambda N, Lambda Lambda and Xi N systems and hypernuclei,''
  Nucl.\ Phys.\ A {\bf 642}, 483 (1998).
  %doi:10.1016/S0375-9474(98)00485-0
  %%CITATION = doi:10.1016/S0375-9474(98)00485-0;%%
  %20 citations counted in INSPIRE as of 06 Nov 2016


%\cite{Batty:1994yx}
\bibitem{Batty:1994yx}
  C.~J.~Batty, E.~Friedman and A.~Gal,
  %``Density dependence of the Sigma nucleus optical potential derived from Sigma- atom data,''
  Phys.\ Lett.\ B {\bf 335}, 273 (1994).
  %doi:10.1016/0370-2693(94)90349-2
  %%CITATION = doi:10.1016/0370-2693(94)90349-2;%%
  %63 citations counted in INSPIRE as of 06 Nov 2016

%\cite{Mares:1995bm}
\bibitem{Mares:1995bm}
  J.~Mares, E.~Friedman, A.~Gal and B.~K.~Jennings,
  %``Constraints on Sigma nucleus dynamics from Dirac phenomenology of Sigma- atoms,''
  Nucl.\ Phys.\ A {\bf 594}, 311 (1995)
  %doi:10.1016/0375-9474(95)00358-8
  [nucl-th/9505003].
  %%CITATION = doi:10.1016/0375-9474(95)00358-8;%%
  %110 citations counted in INSPIRE as of 06 Nov 2016

%\cite{Bart:1999uh}
\bibitem{Bart:1999uh}
  S.~Bart {\it et al.},
  %``Sigma hyperons in the nucleus,''
  Phys.\ Rev.\ Lett.\  {\bf 83}, 5238 (1999).
  %doi:10.1103/PhysRevLett.83.5238
  %%CITATION = doi:10.1103/PhysRevLett.83.5238;%%
  %80 citations counted in INSPIRE as of 06 Nov 2016

%\cite{Noumi:2001tx}
\bibitem{Noumi:2001tx}
  H.~Noumi {\it et al.},
  %``Sigma nucleus potential in A = 28,''
  Phys.\ Rev.\ Lett.\  {\bf 89}, 072301 (2002)
  Erratum: [Phys.\ Rev.\ Lett.\  {\bf 90}, 049902 (2003)].
  %doi:10.1103/PhysRevLett.89.072301
  %%CITATION = doi:10.1103/PhysRevLett.89.072301;%%
  %90 citations counted in INSPIRE as of 06 Nov 2016

%\cite{Saha:2004ha}
\bibitem{Saha:2004ha}
  P.~K.~Saha {\it et al.},
  %``Study of the Sigma-nucleus potential by the (pi-, K+) reaction on medium-to-heavy nuclear targets,''
  Phys.\ Rev.\ C {\bf 70}, 044613 (2004)
  %doi:10.1103/PhysRevC.70.044613
  [nucl-ex/0405031].
  %%CITATION = doi:10.1103/PhysRevC.70.044613;%%
  %82 citations counted in INSPIRE as of 06 Nov 2016

%\cite{Kohno:2006iq}
\bibitem{Kohno:2006iq}
  M.~Kohno, Y.~Fujiwara, Y.~Watanabe, K.~Ogata and M.~Kawai,
  %``Semiclassical distorted wave model analysis of the (pi-,K+) Sigma formation inclusive spectrum,''
  Phys.\ Rev.\ C {\bf 74}, 064613 (2006)
  %doi:10.1103/PhysRevC.74.064613
  [nucl-th/0611080].
  %%CITATION = doi:10.1103/PhysRevC.74.064613;%%
  %37 citations counted in INSPIRE as of 06 Nov 2016

%\cite{Dabrowski:2008zza}
\bibitem{Dabrowski:2008zza}
  J.~Dabrowski and J.~Rozynek,
  %``The (pi-, K+) reaction on Si-28 and the Sigma-nucleus potential,''
  Phys.\ Rev.\ C {\bf 78}, 037601 (2008).
  %doi:10.1103/PhysRevC.78.037601
  %%CITATION = doi:10.1103/PhysRevC.78.037601;%%
  %6 citations counted in INSPIRE as of 06 Nov 2016

%\cite{Haidenbauer:2014uua}
\bibitem{Haidenbauer:2014uua}
  J.~Haidenbauer and U.-G.~Mei\ss ner,
  %``A study of hyperons in nuclear matter based on chiral effective field theory,''
  Nucl.\ Phys.\ A {\bf 936}, 29 (2015)
  %doi:10.1016/j.nuclphysa.2015.01.005
  [arXiv:1411.3114 [nucl-th]].
  %%CITATION = doi:10.1016/j.nuclphysa.2015.01.005;%%
  %10 citations counted in INSPIRE as of 14 Nov 2016

%%\cite{Geng:2013xn}
%\bibitem{Geng:2013xn}
%  L.~Geng,
%  %``Recent developments in SU(3) covariant baryon chiral perturbation theory,''
%  Front.\ Phys.\ (Beijing) {\bf 8}, 328 (2013)
%  %doi:10.1007/s11467-013-0327-7
%  [arXiv:1301.6815 [nucl-th]].
%  %%CITATION = doi:10.1007/s11467-013-0327-7;%%
%  %21 citations counted in INSPIRE as of 23 May 2017

%\cite{Ahn:2005gb}
\bibitem{Ahn:2005gb}
  J.~K.~Ahn {\it et al.} [KEK-PS E289 Collaboration],
  %``Sigma+p elastic scattering cross sections in the region of 350 <= P(Sigma+) <= 750-MeV/c with a scintillating fiber active target,''
  Nucl.\ Phys.\ A {\bf 761}, 41 (2005).
  %doi:10.1016/j.nuclphysa.2005.07.004
  %%CITATION = doi:10.1016/j.nuclphysa.2005.07.004;%%
  %20 citations counted in INSPIRE as of 19 Sep 2017

%\cite{Ahn:1997wa}
\bibitem{Ahn:1997wa}
  J.~K.~Ahn {\it et al.} [KEK-PS E-251 Collaboration],
  %``A Study of Sigma+ p elastic scattering in the region of 300-MeV/c <= p(Sigma) <= 600-MeV/c with a scintillating fiber target,''
  Nucl.\ Phys.\ A {\bf 648}, 263 (1999).
  %doi:10.1016/S0375-9474(99)00028-7
  %%CITATION = doi:10.1016/S0375-9474(99)00028-7;%%
  %24 citations counted in INSPIRE as of 19 Sep 2017

%\cite{Kohno:1999nz}
\bibitem{Kohno:1999nz}
  M.~Kohno, Y.~Fujiwara, T.~Fujita, C.~Nakamoto and Y.~Suzuki,
  %``Hyperon single particle potentials calculated from SU(6) quark model baryon baryon interactions,''
  Nucl.\ Phys.\ A {\bf 674}, 229 (2000)
  %doi:10.1016/S0375-9474(00)00164-0
  [nucl-th/9912059].
  %%CITATION = doi:10.1016/S0375-9474(00)00164-0;%%
  %43 citations counted in INSPIRE as of 13 Nov 2016























%%%%%%%%%%%%%%%%%%%%%%%%%%%%%%%%%%%%%%%%%%%%%%%%%%%%%%%%%%%%%%%%%%%%%%%%%%%%%%%%%%%%
\end{thebibliography}
\end{document}